\documentclass[12pt,preprint]{aastex}

\newcommand{\gsim}{\mbox{\small$\stackrel{>}{\sim}$\normalsize}}	% greater than or similar to

\shorttitle{Metallicities of Open Clusters}
\shortauthors{Marshall et al.}

\received{2005 May 16}
\begin{document}

\title{Survey for Transiting Extrasolar Planets in Stellar Systems. II. 
Spectrophotometry and Metallicities of Open Clusters}

\author{J. L. Marshall, Christopher J. Burke, D. L. DePoy, Andrew Gould, \& Juna A. Kollmeier}
\affil{Department of Astronomy, The Ohio State University}
\affil{140 West 18$^{th}$ Avenue, Columbus, Ohio 43210-1173\\
marshall@astronomy.ohio-state.edu, cjburke@astronomy.ohio-state.edu, 
depoy@astronomy.ohio-state.edu, gould@astronomy.ohio-state.edu, jak@astronomy.ohio-state.edu}

%---------------------------------------------------------------------------

\begin{abstract}

We present metallicity estimates for seven open clusters based 
on spectrophotometric indices from moderate-resolution spectroscopy.  
Observations of field giants of known metallicity 
provide a correlation between the spectroscopic indices and the 
metallicity of open cluster giants. 
We use $\chi^2$ analysis to fit the relation of spectrophotometric indices 
to metallicity in field giants.
The resulting function allows an estimate of the target-cluster giants' 
metallicities with an error in the method of $\pm$0.08 dex.  
We derive the following metallicities for the seven open clusters: 
NGC 1245, [m/H]=$-0.14\pm0.04$; 
NGC 2099, [m/H]=$+0.05\pm0.05$; 
NGC 2324, [m/H]=$-0.06\pm$0.04; 
NGC 2539, [m/H]=$-0.04\pm$0.03; 
NGC 2682 (M67), [m/H]=$-0.05\pm$0.02; 
NGC 6705, [m/H]=+0.14$\pm$0.08; 
NGC 6819, [m/H]=-0.07$\pm$0.12.  
These metallicity estimates will be useful in planning future extra-solar planet
transit searches since planets may form more readily in metal-rich environments.  

\end{abstract}

%---------------------------------------------------------------------

\keywords{Galaxy: Open Clusters and Associations: General, Galaxy: Abundances, Galaxy: Kinematics and Dynamics, Stars: Abundances}

%---------------------------------------------------------------------

\section{Introduction}\label{sec:intro}

Galactic open clusters occupy a wide range of 
ages, sizes, locations in the Galaxy, initial mass functions, and metallicities, 
and are found throughout the Galactic disk. 
Thus, open clusters are excellent tracers of Galactic disk properties
and give insight into the formation of the disk \citep[][]{ja82, jp94, friel95, dlfm2}.
Unfortunately, the fundamental properties of open clusters 
are difficult to determine.  Current
estimates of these properties vary widely; 
there is only marginal agreement in the literature about the age, metallicity, reddening, 
or Galactocentric distance of any given open cluster. 
Part of the disparity between estimates is due to the variety 
of methods used to determine these quantities.  
The published metallicity estimates of an open cluster are especially 
prone to disagreement because each technique has different sensitivities 
to metallicity patterns,
and it is likely that none of the methods, with the possible exception of 
high-resolution spectroscopy, commonly used 
to estimate stellar metallicity is measuring the ``true''
 metallicity of the stars.  
Additionally, the relationship between the many different 
techniques is often unknown.  
The past few decades have yielded significant advances in the 
study of open clusters, but 
there is still much work to be done toward determining the
fundamental properties of the more than 1500 currently known Galactic open clusters.  

Nonetheless, open clusters are extremely useful objects to study because each cluster represents 
a homogeneous set of stars.  
All stars in an open cluster form at the same time and in the 
same circumstances, and thus are expected to have the same age, metallicity, 
and Galactocentric distance.  
For this reason, open clusters are good testbeds for many types of Galactic studies.  
For example, they may be ideal objects in which to search for extra-solar planets
\citep[e.g.,][]{gonzalez97}.   
Furthermore, open clusters can be metal-rich objects. 
Metal-rich environments have recently received attention 
due to the suggestion \citep[e.g.,][]{santos, gonzalez} that extra-solar planets 
form more readily around metal-rich stars.  If this is the case, metal-rich 
open clusters may be excellent places to search for extra-solar 
planets.  This is the primary motivation for this work; planet searches will be 
conducted in metal-rich open clusters identified in this study.  These programs 
will be carried out by the Survey for Transiting Extrasolar Planets 
in Stellar Systems (STEPSS) program \citep[e.g.][]{burke03}.  

In this paper we present new, accurate metallicity estimates for seven 
open clusters that were presumed to be metal-rich based on literature searches.  
We have obtained moderate resolution spectroscopy of 
giants in the target-clusters and use spectrophotometric line indices to 
produce metallicity estimates for the target-clusters.  

%---------------------------------------------------------------------------

\section{Observational Procedure}\label{sec:obs}

In this work we follow the method of \cite{friel87} to determine metallicities 
using spectrophotometric line indices as proxies for 
high-resolution metallicity estimates.
This procedure was adopted in several subsequent papers, including 
\cite{tff}, \cite{fj93}, and most recently \cite{friel02}.  
Here we quantitatively assess the precision of this methodology, 
which is described below.  

\subsection{Sample Selection}\label{sec:sample}

The target-clusters were selected as a precursor to a search 
for extra-solar planets.  Clusters to be observed were 
chosen on the basis of a high
metallicity estimate cited in the literature, as well as their location in the Galaxy.  
We used the \cite{lynga} database
to select target-clusters, 
with selection criteria of [Fe/H] $> -0.2$ so as not to exclude 
metal-rich clusters with large errors in their metallicity determination.  
We required the target-clusters to have 
distances $>$ 300 pc, so that the stars would be bright enough to observe on our 2-m class telescopes,
and to have declinations $\delta$ (1950) $> -15^o$, so as to be 
observable from the northern hemisphere. 
We did not put any constraints on the ages of the clusters; indeed, the 
selected open clusters span a wide range of ages.
These criteria yielded a set of 26 candidate open clusters.

These target-clusters were observed with the MDM ``Echelle'' CCD imager  
at the 1.3 m McGraw-Hill telescope of the 
MDM Observatory\footnote[1]{See http://www.astro.lsa.umich.edu/obs/mdm/} in a preliminary observing run.  
During this run we obtained images of the 
target-clusters as well as instrumental color-magnitude diagrams 
(CMDs) of each cluster.  
We use these images and instrumental CMDs to select target-cluster stars.

Using these preparatory images, 
we further required the selected open clusters to be optically 
distinguishable as a cluster in the 1.3 m images (the imager has a 17\arcmin field of view). 
This constraint ensures a large number of observable stars, both in the clump 
(for this work) and in the cluster in general (for use in the planet searches). 
The clusters were also explicitly required to have several giant stars in or near the clump,
based on the instrumental CMD, in order to determine accurately the metallicity of each target-cluster.  
The seven target-clusters selected from these criteria are 
NGC 1245, NGC 2099, NGC 2324, NGC 2539, NGC 2682, NGC 6705, 
and NGC 6819.  
We select giants with $(B-V)_0\approx1$ 
(typically red clump giants) in each of the target-clusters. 
For reference, the names of the stars in each target-cluster used here are those given
in the WEBDA open cluster database\footnote[2]{See http://obswww.unige.ch/webda/}.
We also give the references for the identifications for each cluster in \S \ref{sec:metal}

We also observed stars of known metallicity to use as calibrators.  
We took calibration stars mainly from the survey of \cite{friel87}, who presents a large
sample of bright field stars and cluster giants with accompanying 
measured indices and previously derived metallicities.  
We took bright metal-poor calibration stars from \cite{cs86}, 
while additional giant stars from the old open cluster NGC 2682 were taken from \cite{fj93}.  
In order to minimize calibration errors between target-cluster stars and 
the calibrators, we selected calibration stars from these papers 
that have a color of $B-V=1 \pm 0.1$ mag, roughly
the same (dereddened) color as the target-cluster stars.  
Additionally, we observed a few calibration 
stars with $B-V<0.9$, in order to ensure some overlap 
with the color of the target-cluster stars.  

\subsection{Observations and Data Reduction}\label{sec:data}

The spectral data were obtained at the MDM Observatory using the CCDS 
spectrograph\footnote[3]{See http://www.astronomy.ohio-state.edu/MDM/CCDS/}.
We obtained one night of data (2002 January 31) 
on the 2.4 m Hiltner telescope, while the other seven nights of 
data (during 2002 March 21 - April 01) were obtained on the 1.3 m 
McGraw-Hill telescope.  The spectra cover the wavelength range 
3800-5400 \AA \ with the 350 l/mm (1.33 \AA/pix) grating.  
The full-width at half-maximum (FWHM) of an unresolved emission line is $\sim$3 pixels, 
yielding a spectral resolution ($\lambda$/$\Delta\lambda)$ of $\sim$1150.  
All spectra on both telescopes use 
an 87 $\micron$ slit for uniformity.  This slit size translates to 
1\arcsec \ on the 2.4 m and 1\farcs 8 on the 1.3 m.  Most spectra have 
formal signal-to-noise ratios (S/N) \gsim 100 per pixel.  

\begin{figure}
\plotone{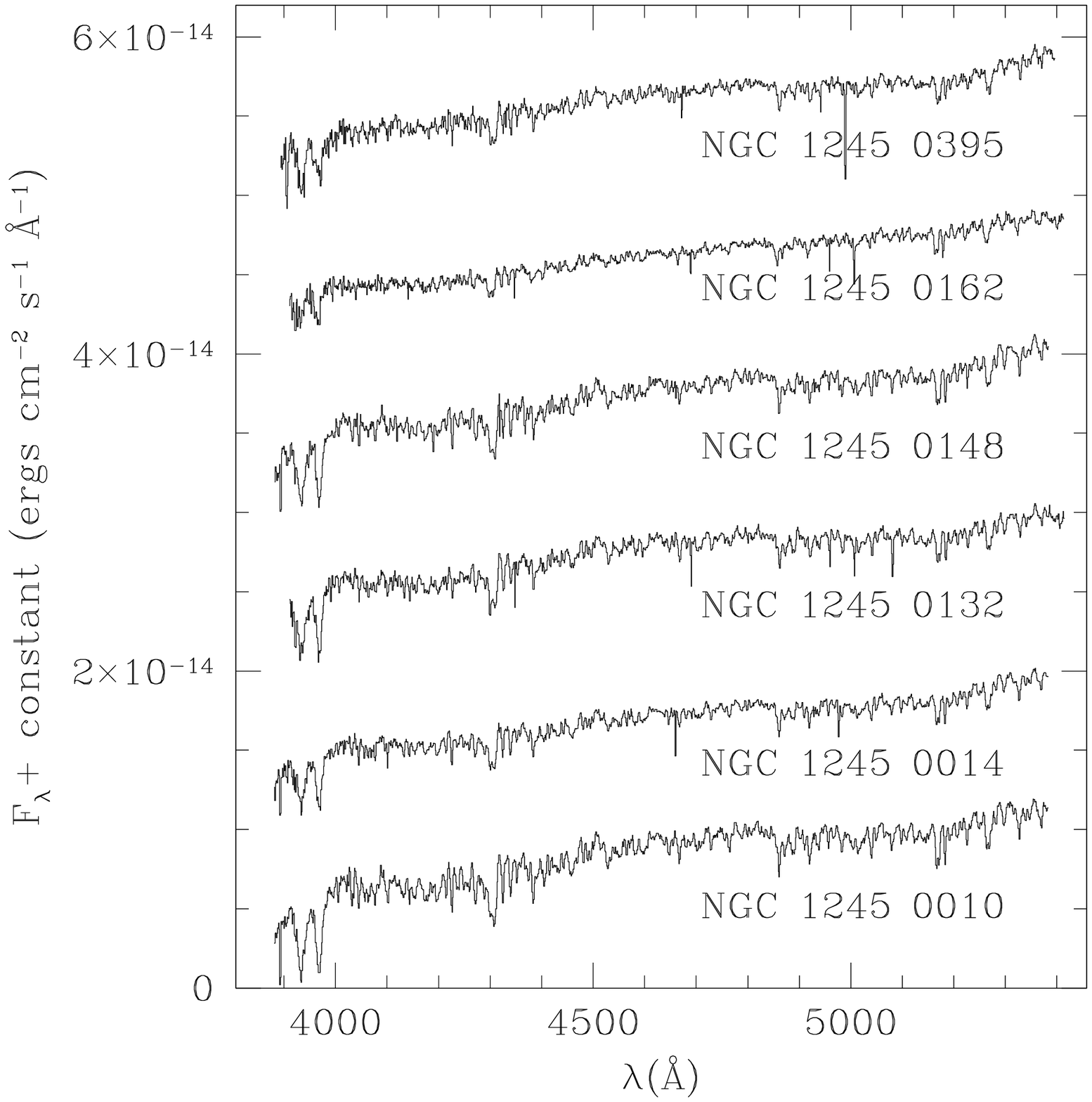}
\figcaption{
Spectra of all target-cluster stars observed in the cluster NGC 1245.  Each spectrum includes an additive constant for display purposes.
\label{fig:f1}
}
\end{figure}

\begin{figure}
\plotone{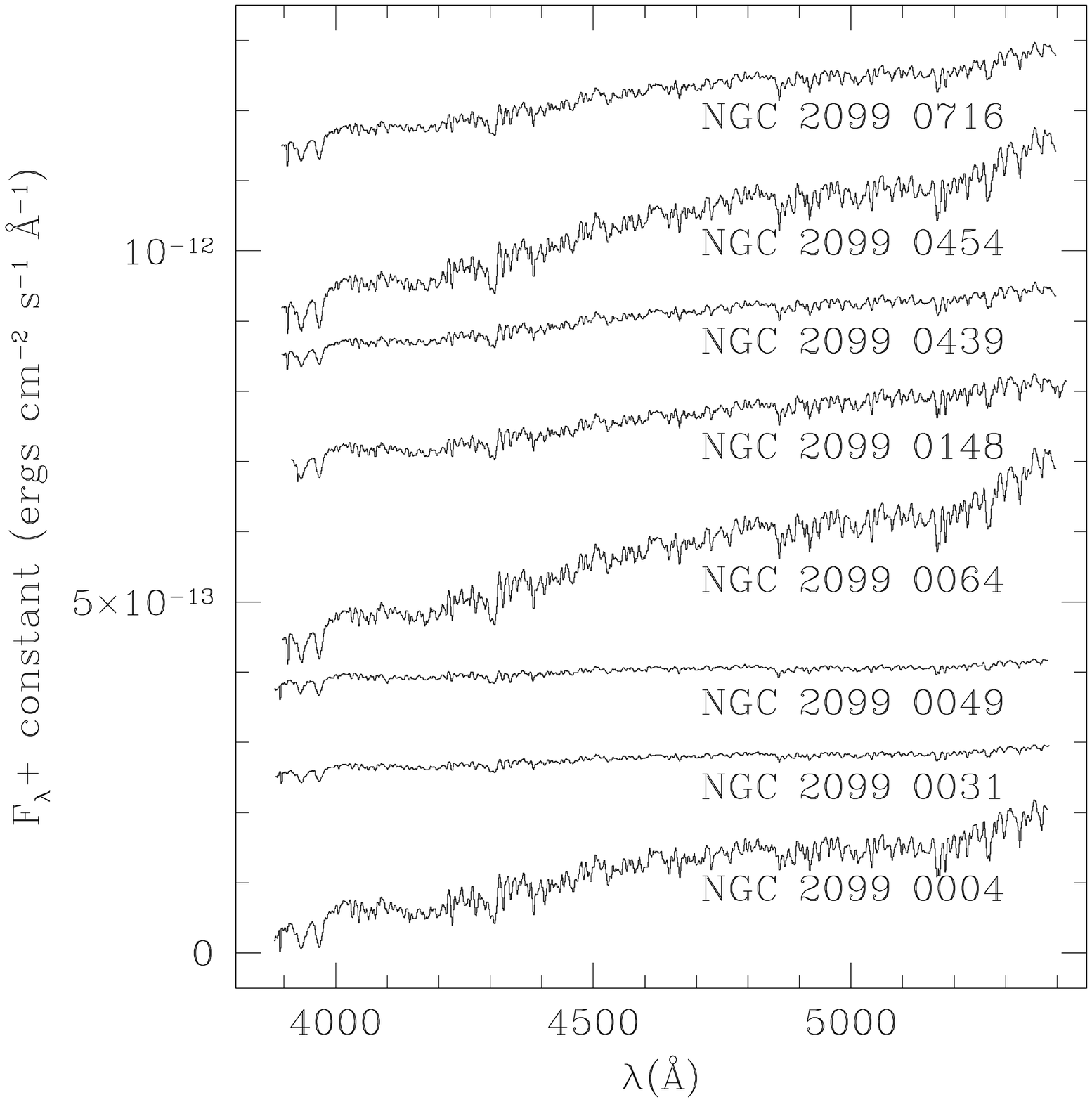}
\figcaption{
Same as Figure \ref{fig:f1}, but for the cluster NGC 2099.
\label{fig:f2}
}
\end{figure}

\begin{figure}
\plotone{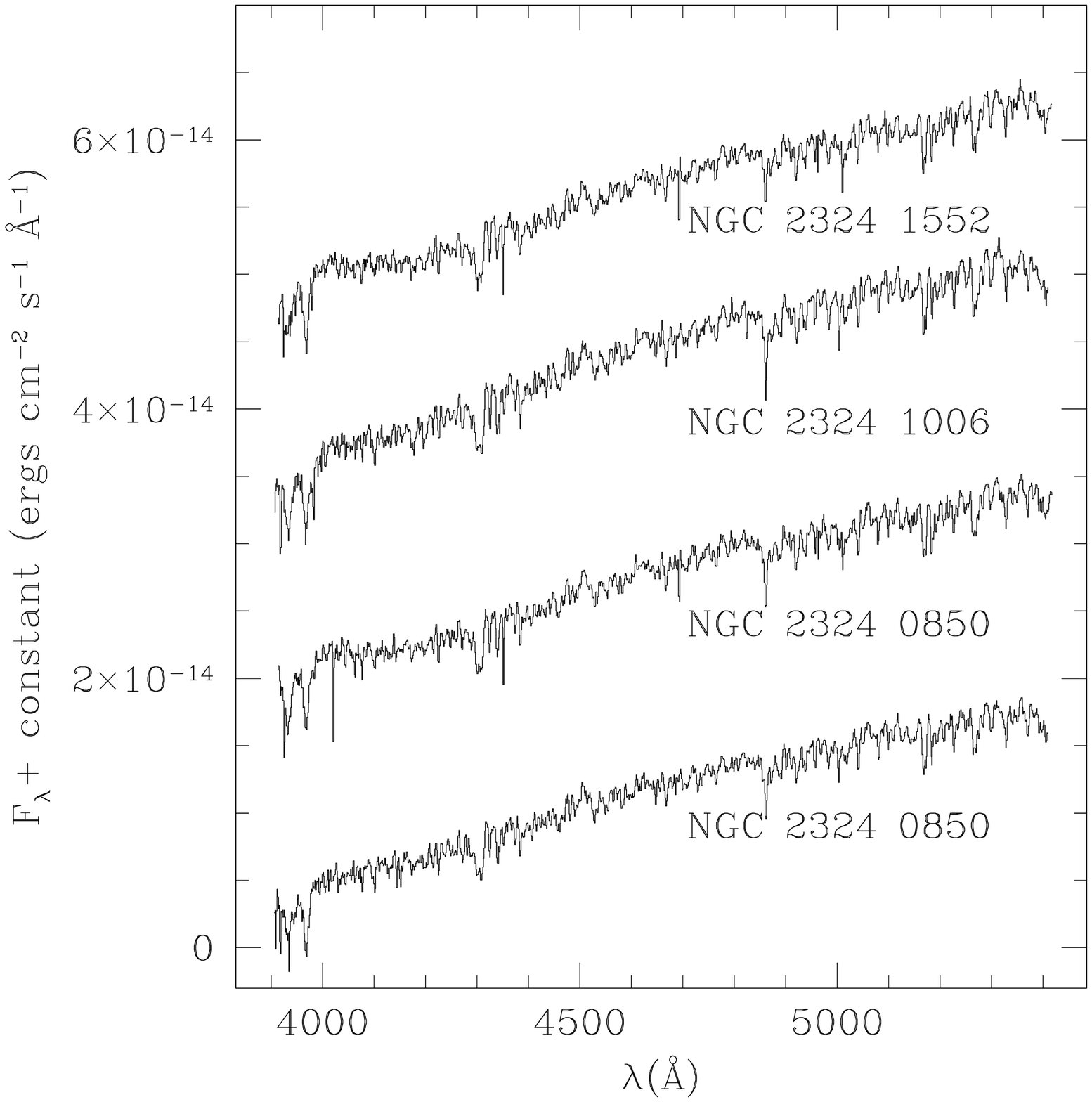}
\figcaption{
Same as Figure \ref{fig:f1}, but for the cluster NGC 2324.
\label{fig:f3}
}
\end{figure}

\begin{figure}
\plotone{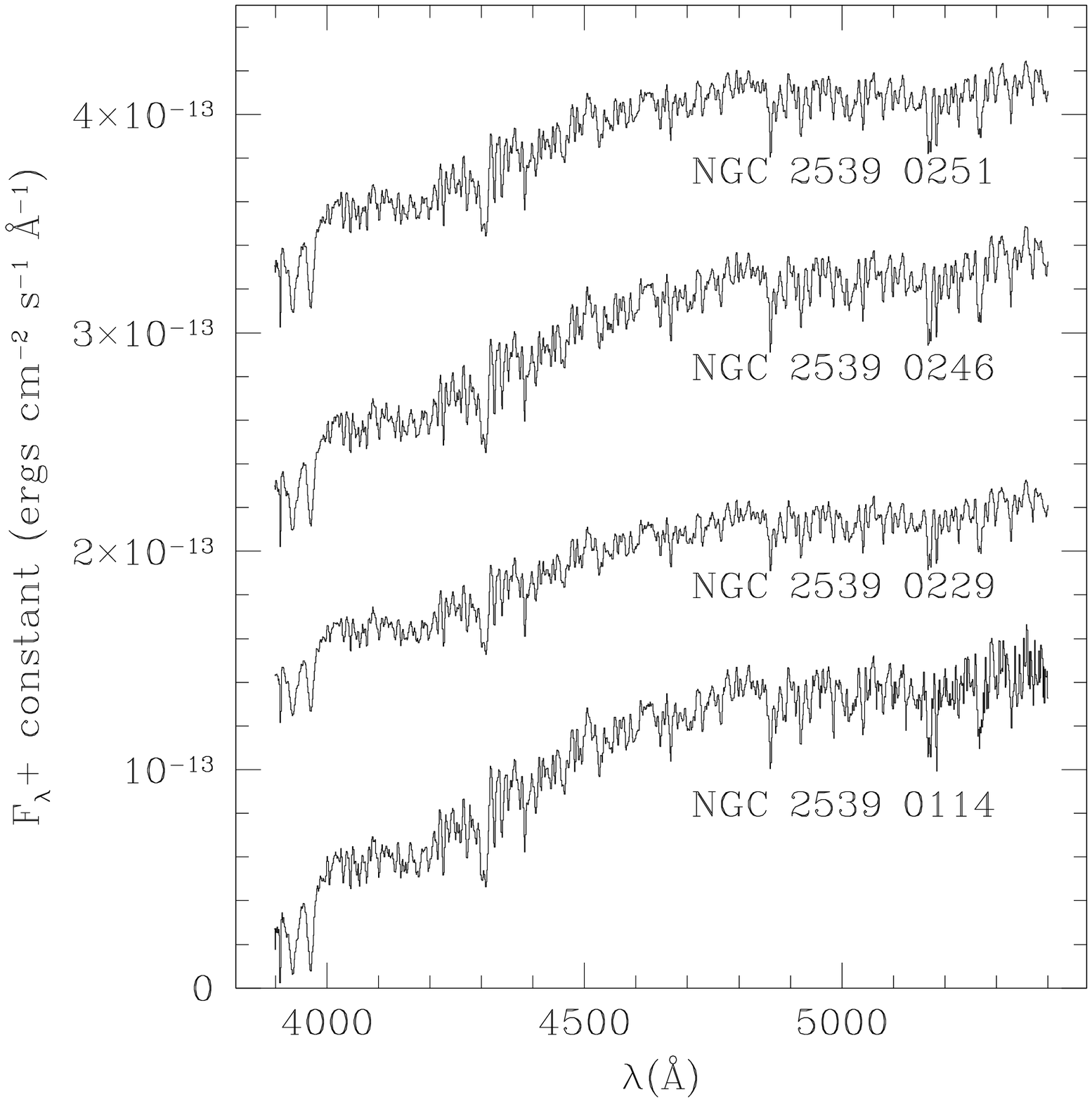}
\figcaption{
Same as Figure \ref{fig:f1}, but for the cluster NGC 2539.
\label{fig:f4}
}
\end{figure}

\begin{figure}
\plotone{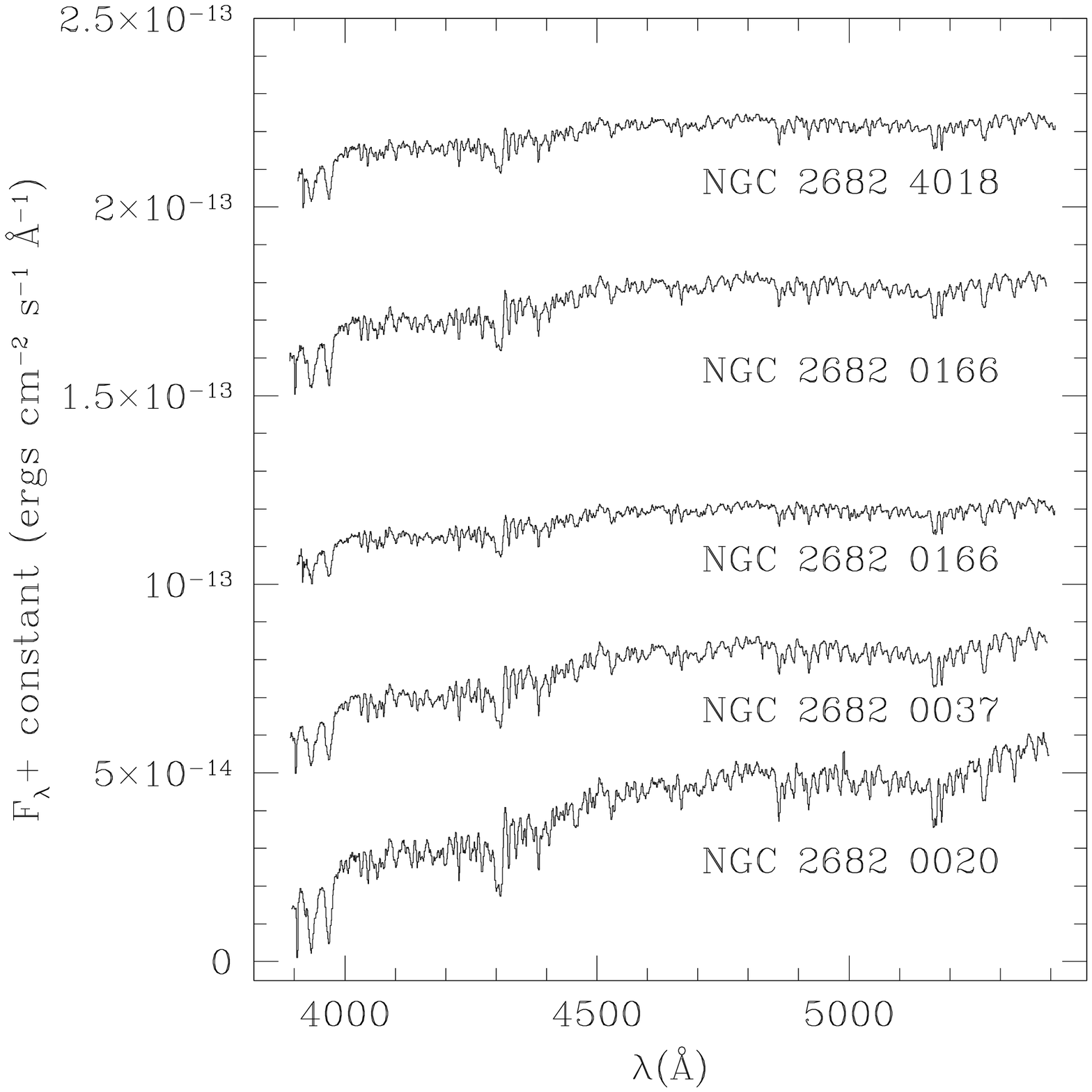}
\figcaption{
Same as Figure \ref{fig:f1}, but for the cluster NGC 2682.
\label{fig:f5}
}
\end{figure}

\begin{figure}
\plotone{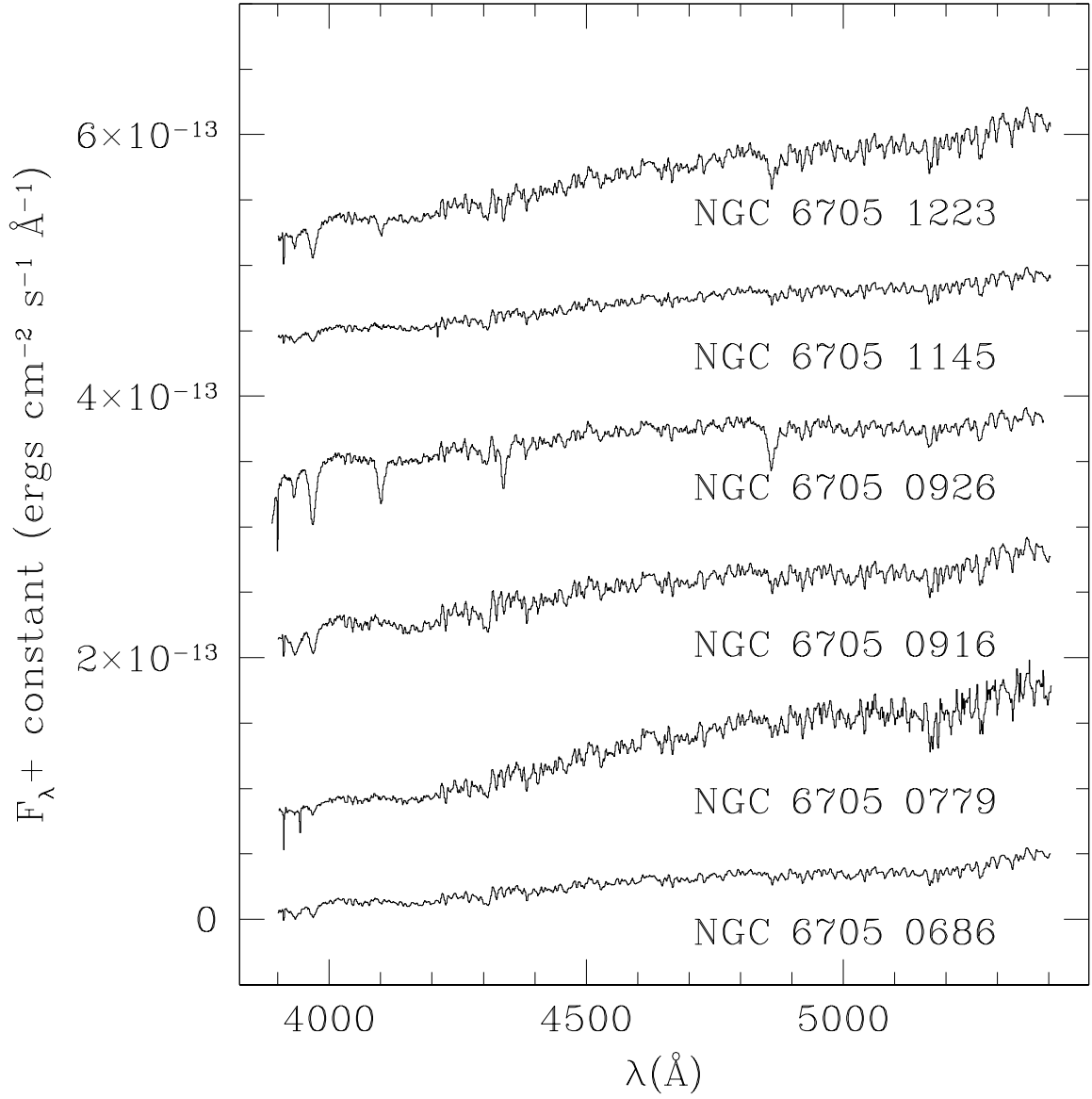}
\figcaption{
Same as Figure \ref{fig:f1}, but for the cluster NGC 6705.
\label{fig:f6}
}
\end{figure}

\begin{figure}
\plotone{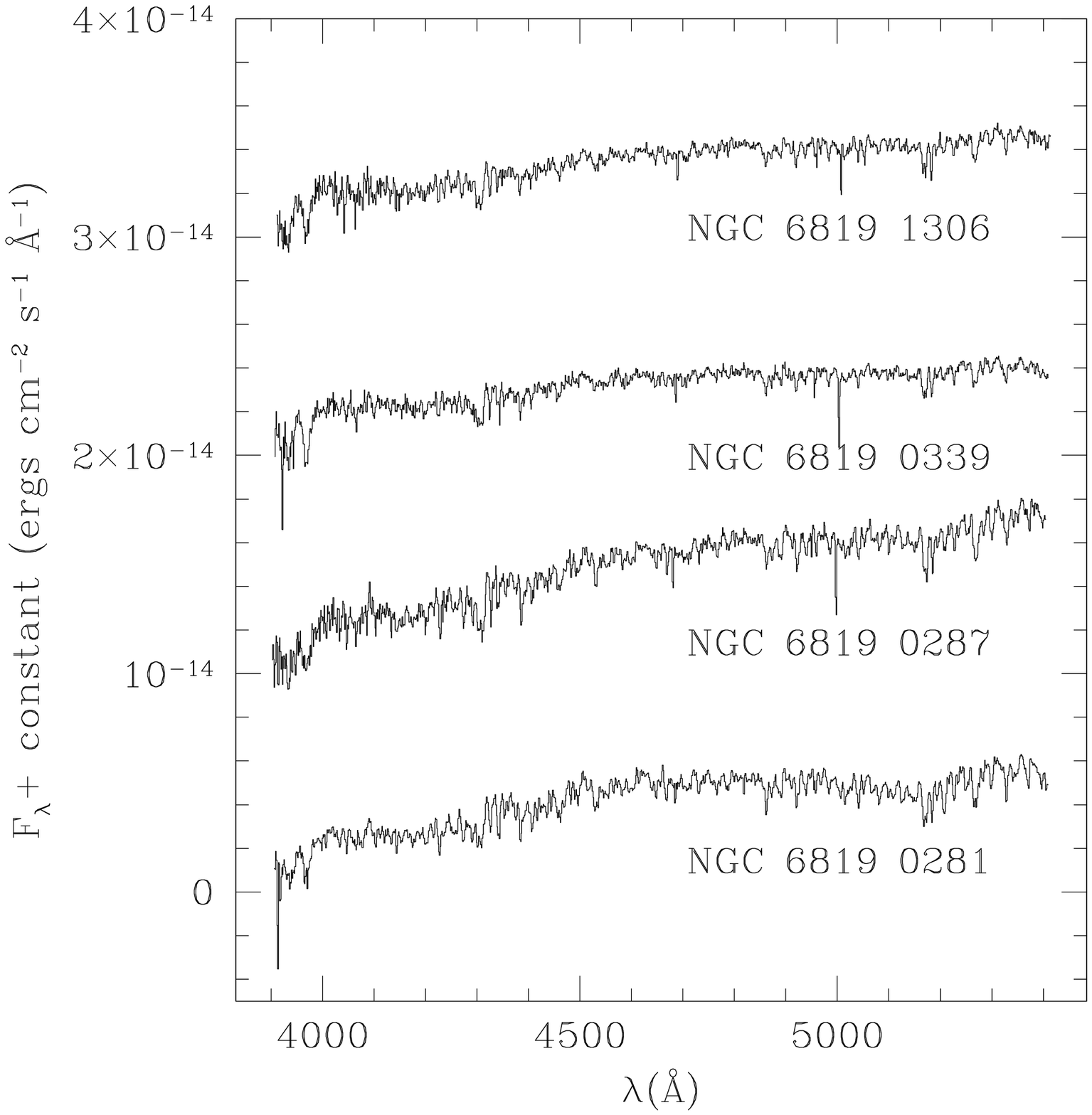}
\figcaption{
Same as Figure \ref{fig:f1}, but for the cluster NGC 6819.
\label{fig:f7}
}
\end{figure}

\begin{figure}
\plotone{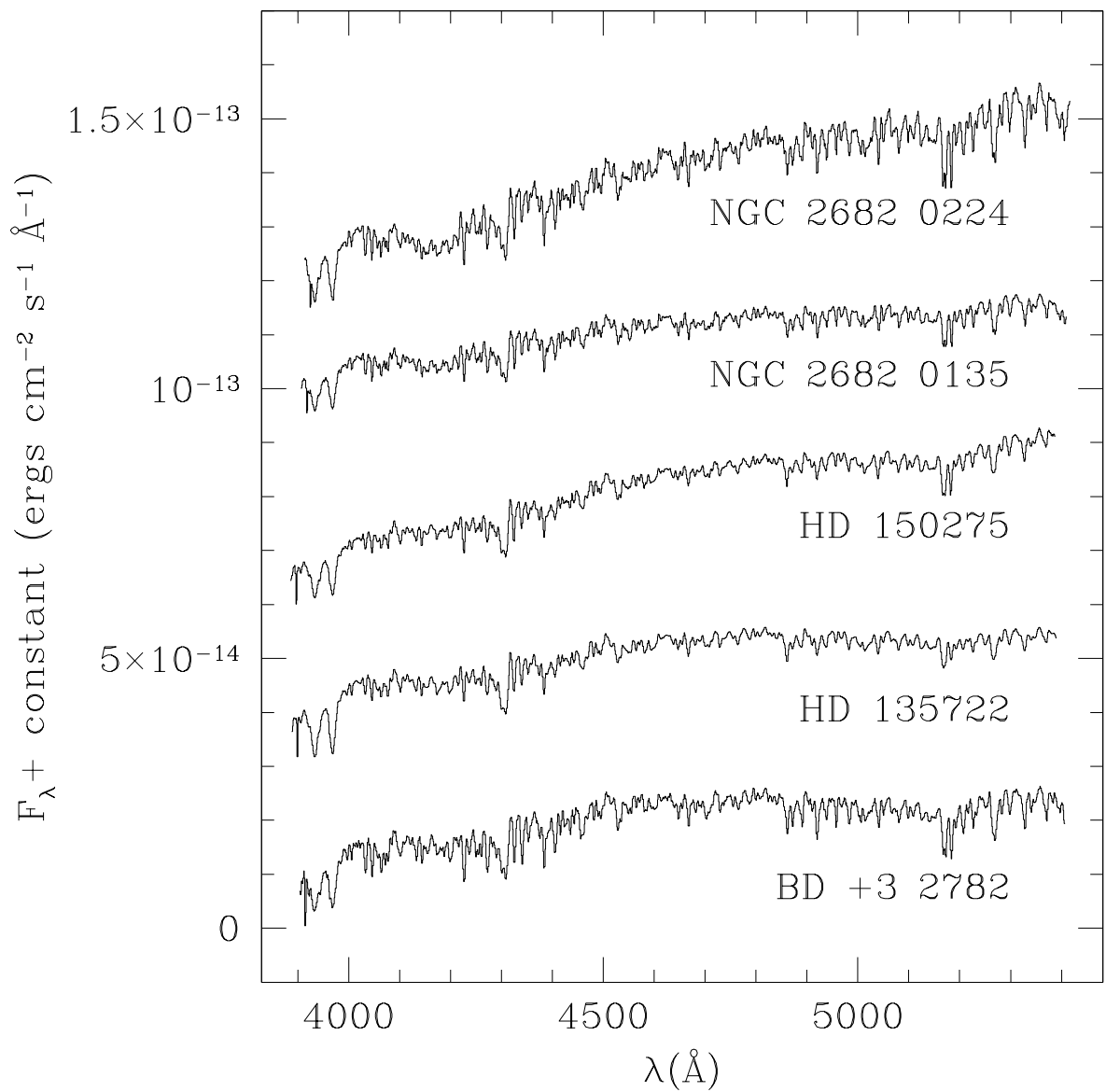}
\figcaption{
Spectra of selected calibration field red giants in a range of colors.  
Shown are BD +3 2782 ($B-V$=0.9), HD 135722 ($B-V$=0.95), HD 150275 ($B-V$=1.0), NGC 2682 0135 ($(B-V)_0$=1.06), and NGC 2682 0224 ($(B-V)_0$=1.1).
The spectra are normalized to BD +3 2782 at 4500 \AA, and include an additive constant for display purposes.  
\label{fig:f8}
}
\end{figure}

We observed the target-cluster stars interspersed with calibration stars of known metallicity.  
Thirty-five observations of 33 target-cluster stars and
56 observations of 38 calibration stars were made.
Figures \ref{fig:f1} -- \ref{fig:f7} show the reduced spectra 
of all of the stars observed in each target-cluster.  
Figure \ref{fig:f8} shows a selected sample of the calibration stars observed.
Multiple observations of some stars were included in order 
to ensure reproduction of Friel's method, to check repeatability 
of the measurements between telescopes, and also to have many 
measurements on which to base the calibration.  

We reduced the data using standard IRAF\footnote[4]{IRAF is 
distributed by the National Optical Astronomy Observatory, which is 
operated by the Association of Universities for Research in Astronomy, 
Inc., under cooperative agreement with the National Science Foundation.} 
routines following the usual 
practice of overscan subtraction, division by a flat field, and extraction of 
the spectra.  Flat fields and comparison lamp spectra 
were obtained using lamps within the instrument.  
We correct the spectra for atmospheric extinction using the standard KPNO (Kitt Peak) extinction 
curves provided by IRAF.  

On each night we observed a spectrophotometric standard star from \cite{strom}.
The application of only one standard per night (regardless of the airmass of the 
observations) may introduce a small color effect
into the spectra since the airmass has a color term and we do nothing 
to account for the airmass of the observations.  Due to the nature of the 
measurements (indices with continuum bands on either side of the lines), 
this is not a significant concern.
No attempt was made to do spectrophotometry, i.e., we did not align the slit 
with the parallactic angle at each telescope pointing.  
This seems to have had little effect on our results.  

\subsection{Measurement of Spectrophotometric Indices}\label{sec:measure}

Several recent papers \citep[e.g.][]{sb04} use spectrophotometric indices as an 
estimator for the
metallicity and other properties of individual stars.  The indices are measured 
for both the program stars as well as calibration stars with well-known 
metallicity; the calibration star data are fit to obtain a 
calibration for the metallicities of the target-cluster stars.  This practice is desirable 
because the indices may be measured with moderate resolution spectra to 
obtain a relatively accurate (i.e., generally good to $\sim$ 0.15 dex) metallicity estimate for the star, and 
obviates the need for difficult-to-obtain high-resolution spectroscopy.  
Lick indices \citep[][]{burstein} are commonly used in this technique and are applied 
to extragalactic sources for which it is often impossible to obtain high 
S/N, high-resolution spectra.  \cite{friel87} slightly 
altered the Lick indices to account for differences between Galactic and 
extragalactic sources and applied them to giant stars in the Galaxy.  

The indices are defined as a ratio between a bandpass that incorporates several strong 
metal lines and continuum measurements on either 
side of this bandpass that include little flux from metal lines.  The final 
measured indices (several per star) may then be combined to yield 
a metallicity estimate for each star.  
For example, the Fe4680 index is defined as 

\begin{equation}
{\rm Fe}4680 = -2.5 log \frac{ \int_{4636.00}^{4723.00} F _\lambda d\lambda / 87 \AA }{ \int_{4606.00}^{4636.00} F _\lambda d\lambda / 30 \AA  +  \int_{4736.00}^{4773.00} F _\lambda d\lambda /37 \AA }.
\end{equation}
  
The line and continuum regions of the indices used in this work are given in Table \ref{table:table1}, 
as well as the species measured by each index.  This table is reproduced from \cite{friel87}.

\begin{figure}
\plotone{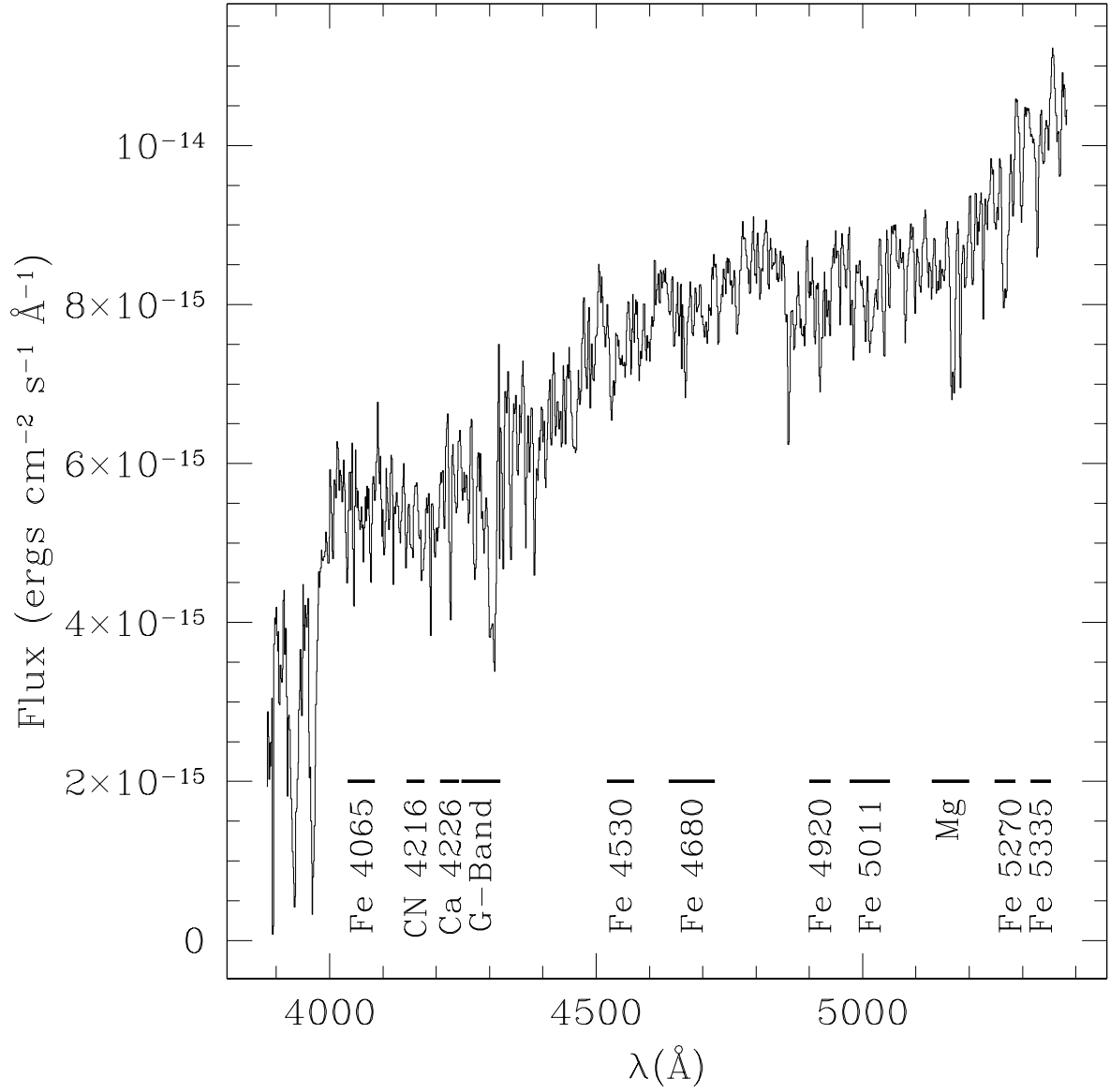}
\figcaption{
An example spectrum (of the target-cluster star NGC 1245 0148) showing the central bandpasses and widths
of indices measured in this paper.
\label{fig:f9}
}
\end{figure}

\begin{figure}
\plotone{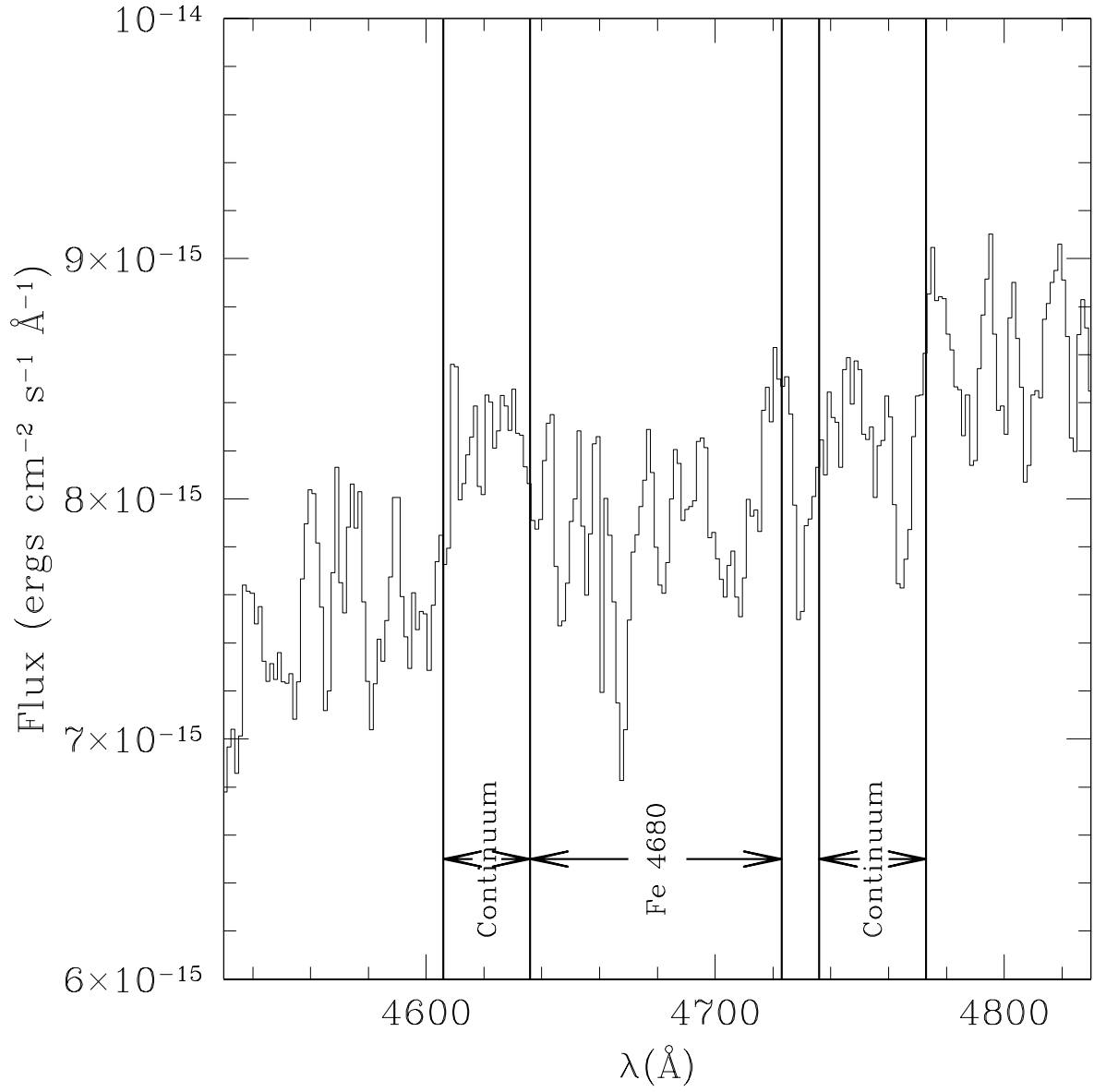}
\figcaption{
An enlarged portion of the spectrum shown in Figure \ref{fig:f9}, detailing the placement of one index (Fe4680) and its flanking continuum bandpasses
\label{fig:f10}
}
\end{figure}

Figure \ref{fig:f9} shows a sample moderate resolution spectrum obtained on the 
MDM 2.4 m telescope.  The central bandpasses of each measured index are 
shown across the bottom of the figure.  Figure \ref{fig:f10} shows an enlarged section 
of the same spectrum, showing the central bandpass and 
accompanying continuum bandpasses for the Fe4680 index.  

%---------------------------------------------------------------------------

\section{Analysis}\label{sec:analysis}

\subsection{Metallicities of Calibration Stars}\label{sec:calib}

Although the calibration stars are selected from \cite{friel87}, \cite{cs86}, and \cite{fj93}, 
many of these stars have more recent metallicity estimates than those cited in these 
original references.  
In order to minimize errors introduced into our analysis from the 
calibration stars, we selected only field giants from \cite{friel87} that had 
metallicity estimates from a single source.  
Specifically, we chose metallicity estimates from the \cite{cds01} 
(hereafter CdS) catalog.  
This catalog is a compilation of previously published atmospheric parameters          
determined from high-resolution, high S/N
spectra from a variety of sources.  The          
2001 version of the catalog contains bright F, G, and K stars in the Galactic field 
as well as in certain associations.         
We selected thirty-eight metallicity standards from \cite{friel87} that had 
at least one metallicity determination in the CdS catalog; 
we obtained 56 measurements of these 38 stars.  

For stars with more than one entry in the CdS catalog we combine the 
available entries as follows.  If there are two entries we average the 
two quoted metallicities.  For stars with 
three or more entries, we discard metallicity measurements more than one 
standard deviation away from the median of all the measurements.  We then 
average the remaining measurements.  When only one measurement was present 
in the catalog, we adopt it.

Many of the entries in CdS do not have associated error estimates; those that do are 
of order 0.1 dex.  We estimate errors by finding the standard deviation of the 
remaining metallicities once spurious measurements are discarded as described above.  
For stars with only one or two measurements, or those for which only two measurements remained, 
we assume an error of 0.1 dex.  
However, this may be an overestimate of these errors, as our method reproduces 
metallicities of most of these field stars to higher precision (see \S \ref{sec:fit}).

\subsection{The Indices: Reproducibility and Comparison to Previous Index Measurements}\label{sec:indices}

Table \ref{table:table2} gives the measured indices of the calibration stars observed in 
this work along with their $V$ mag and $B-V$ colors \citep[from the Hipparcos Catalog,][]{esa97} 
and metallicities (derived from the CdS database, as described above).  
Table \ref{table:table3} shows these data for the target-cluster stars.  

We obtained all of the spectra with the CCDS spectrometer, and 
we made most of the observations with the MDM 1.3 m telescope.  Since 
all measurements were obtained with the same instrument, we do not expect 
large systematic differences between the data obtained on each telescope.  
Several stars were observed on both of the MDM Observatory telescopes.  
We compared the indices for the stars observed on both
telescopes; the results are shown in Table~\ref{table:table4}.  
The average difference between the target-cluster stars measured on the two telescopes is 
0.00 and the standard deviation between the stars measured 
on each telescope is 0.03.  This difference is well within the errors of the 
measurements and suggests no systematic difference in the 
indices measured on the two telescopes.  

We also observed some stars twice on the 1.3 m run.
We compare the indices for 
the stars observed twice on the 1.3 m run, as well as the indices of 
the stars observed multiple times over both runs.  
No stars were observed twice on the 2.4 m run.  
The results are again shown in Table~\ref{table:table4}.  Note that in both cases 
the average difference between any two measurements is 
0.00.  The standard deviation for the stars measured twice 
during the 1.3 m run is 0.04 while that for all stars measured twice 
(over both runs) is 0.03.  

Finally, we compare our measured indices for the calibration stars to 
those given in \cite{friel87}.  
For all of the stars that were measured both by \cite{friel87} and 
in this work, we found the difference between Friel's value of each index 
and ours and computed the average and standard deviation for all of the indices.  
Our indices reproduce Friel's well; the average difference of the 
indices is $-0.012$ and the standard deviation is 0.054.  

\subsection{Reddening}\label{sec:reddening}

In order to determine quantitatively if the interstellar reddening to the target-clusters
affects our results, we simulated 
the effects of reddening on a sample spectrum.  
We used IRAF utilities to redden artificially the spectra and measured the 
effect of reddening on each of the indices.  We applied an E(B-V) of 1.0 
(much more extreme than any of the target-clusters should actually have; see footnotes to Table \ref{table:table3}), 
and re-calculated the index values.  For some indices the difference in the index
between the raw spectrum and the dereddened spectrum was higher than the others
(Ca4226 and G-band were 0.05 and 0.11 respectively).  For the remaining nine 
indices the difference was 0.03 or less (roughly the precision of the index measurements).  
However, none of the target-clusters have nearly this high of a reddening value 
(see the footnotes to Table \ref{table:table3}).  At the approximate reddening values applicable to 
our target-clusters, the effect of reddening was well below the other sources of error.
We conclude that reddening does not significantly affect our results.  

%---------------------------------------------------------------------------
\section{Results}\label{sec:results}

\subsection{Fit to Calibrators}\label{sec:fit}

We begin by writing the predicted metallicity as a linear               
function of the 11 observed spectrophotometric line indices $f_i$ ($i=1\ldots 11$),              
${\rm [Fe/H]}_{\rm pred} = \sum_{i=0}^{11} a_i f_i$, where                      
$f_0=1$.  We determine the coefficients $a_i$ by making a linear                
fit to 56 stars that have spectroscopically determined metallicities from CdS of 
[Fe/H]$>-1.5$.          
                                                                                
We determine which of these 11 spectrophotometric line indices are                         
redundant as follows.  In the fit we arbitrarily assign each             
index an error of unity.  Since there are $56-12=44$ degrees of freedom (dof),        
the resulting $\chi^2/$dof of the fit gives the (square of the)                 
scatter about the fit.  We then try removing each of the 12 parameters          
in succession and find the parameter that causes $\chi^2$ to                    
increase by the least.  Since the number of dof has also increased (by 1),      
the $\chi^2/$dof may have increased or decreased.  If it has                    
decreased, we regard the parameter as redundant, remove it, and then            
try removing each of the remaining 11 parameters.  We continue this             
process until $\chi^2/$dof begins to increase.  

We derive the metallicities of the target-clusters 
using the 56 calibrator stars.  We find that the target-cluster stars all have [Fe/H]$>-0.7$.  
Further, we find that thirteen calibrator stars have extremely high 
scatter in this fit ($\sigma>$0.15), possibly due to large errors in their CdS metallicity determinations.  
We therefore repeat the entire procedure but restricted to the sample of 43 
calibrator star measurements that have [Fe/H]$>-0.7$, 
discarding the high-scatter stars.  In this case,                  
we find 5 indices are removed (Fe4530, Fe5011, Mg, Fe5270, Fe5335).  
The equation used to combine the indices and produce the metallicities of the stars is 

\begin{eqnarray}
\lefteqn{\nonumber[m/H]=} \\
& & -3.180415+4.398434\times\mbox{Fe4065}+1.105911\times\mbox{CN4216} + \\
& & \nonumber-2.480848\times\mbox{Ca4226}+-0.725162\times\mbox{G-band} + \\
& & \nonumber2.835423\times\mbox{Fe4680} +-1.620898\times\mbox{Fe4920}.
\end{eqnarray}
The coefficients of the indices are also presented in Table~\ref{table:table5}.
                                                                                
The scatter about the relation is $\sqrt{\chi^2/\rm dof}=0.0709$.        
This number reflects the quadrature sum of the                                 
root-mean-square (rms) error in the spectroscopic metallicities and the intrinsic                  
scatter of the method.  

Even though the target-cluster giants are selected to have approximately the same $B-V$ color 
as the calibration stars, we find that the derived metallicities of the target-cluster giants 
have a non-negligible dependence on color.  
This effect is due to the fact that some of the stars we observed were not in 
the red giant clump but rather were more evolved stars on the giant branch.
Thus these stars have different surface temperatures and gravities than 
the rest of our clump stars.  This effect is readily calibrated by removing 
the color dependence of the metallicity.  

We compare metallicity residual (relative to CdS metallicity) to $B-V$ color
for the five calibrator stars and five target-cluster stars in NGC 2682 
(the target-cluster with the most well-determined metallicity in the literature).  
The $B-V$ colors of these stars are dereddened using E(B-V)=0.05 \citep[][]{mont}.
For the five program giants (with no literature metallicity values) we assume the 
average literature value for the cluster metallicity of [Fe/H]=$-0.05$.
The equation of the fit is
 
\begin{equation}
[Fe/H]_{CdS} - [m/H]_{this paper} = 1.4937\times(B-V) - 1.5858.
\end{equation}

\begin{figure}
\plotone{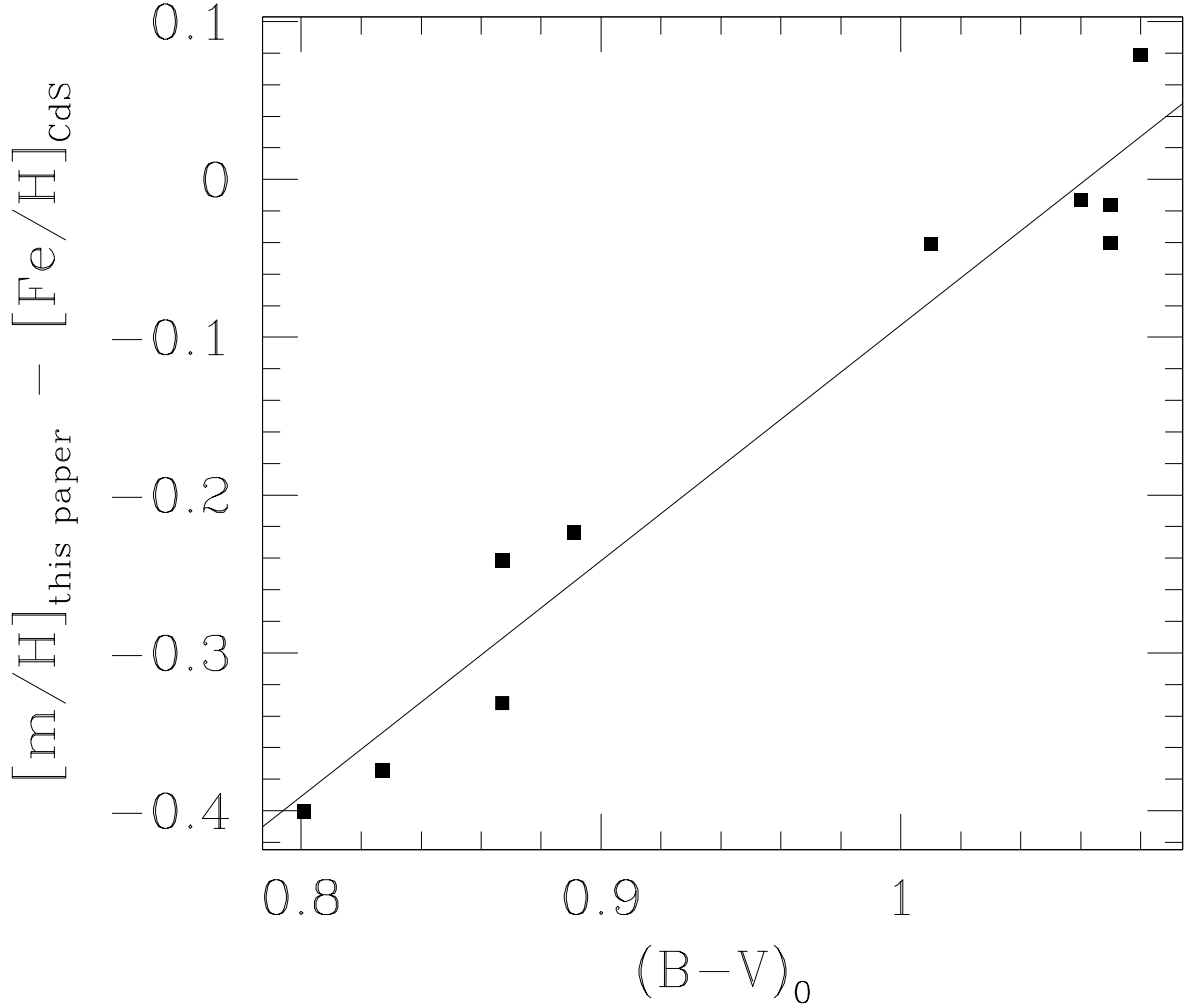}
\figcaption{
Derivation of the color correction term applied to the target-cluster stars to determine their final metallicities.  
\label{fig:f11}
}
\end{figure}

The derived color correction is then applied to 
the 35 target-cluster stars individually
to obtain the final metallicity of each star.
We assume that the spread in ages of the target-clusters has 
a negligible effect on this fit.  More explicitly, we assume the 
errors associated with this assumption are smaller than our other sources of error.  
The data and the derived fit are shown in Figure \ref{fig:f11}.

\begin{figure}
\plotone{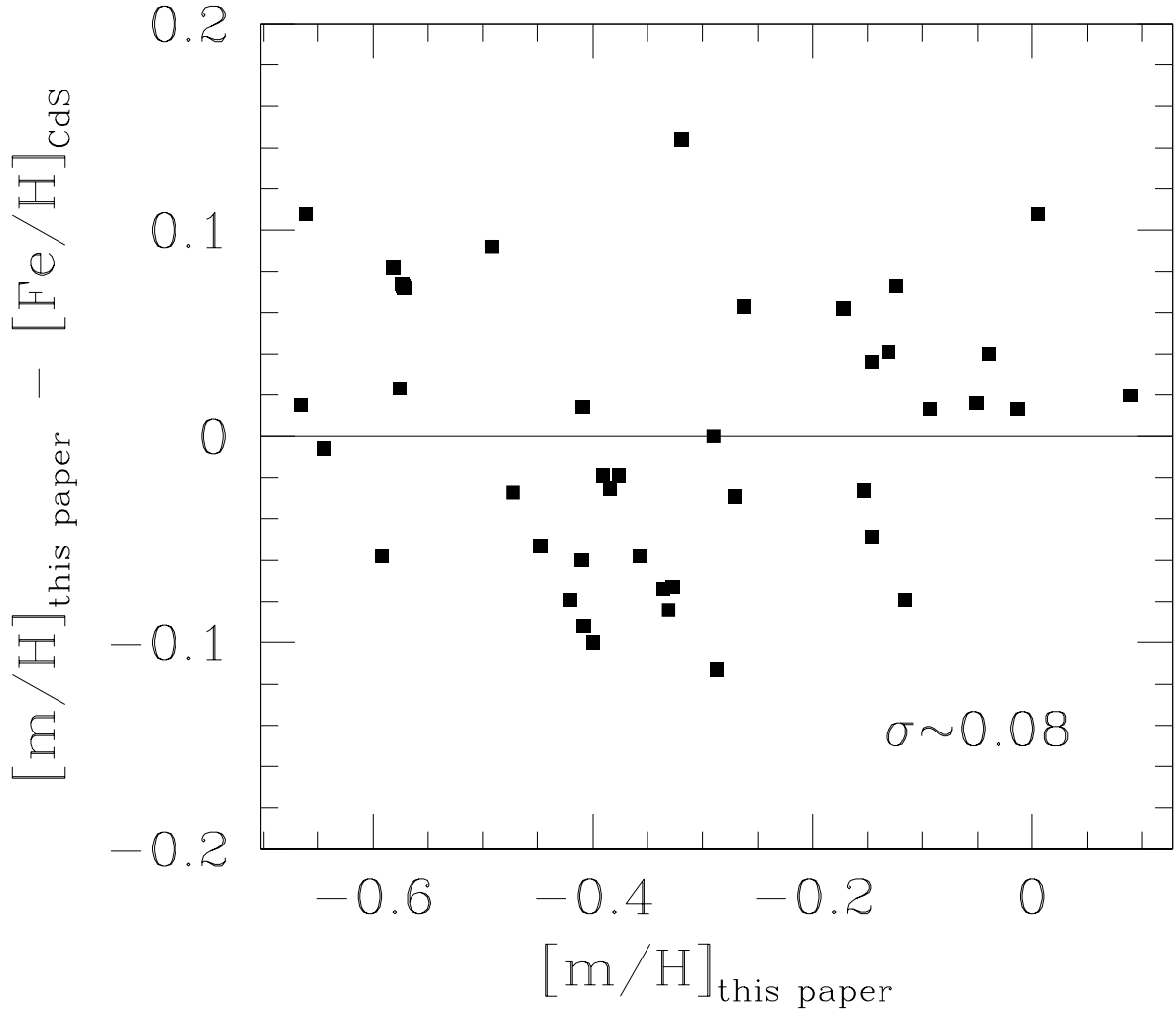}
\figcaption{
Comparison of metallicities derived here with those of \cite{cds01}.  The scatter is $\sim$0.08 dex.
\label{fig:f12}
}
\end{figure}

Figure \ref{fig:f12} compares our derived metallicities 
(after applying the color correction) to 
those of CdS. The scatter in this relation is $\sim$0.08 dex, which is
only 0.01 dex larger than the scatter in the derivation of the original relation.  
We therefore estimate the overall error in the method to be $\sim$0.08 dex.
This implies that the technique of applying a color correction to the 
derived metallicities produces acceptable results.

Note that we use the term [m/H] to refer to our metallicity estimates.  
Since each of the spectrophotometric line indices is selected to incorporate several metal absorption lines,  
our method must measure some aspect of the star's metallicity.
But since the indices comprise several different metal species, 
this method does not necessarily measure only the iron abundance of the star.

\subsection{Cluster Metallicities and Comparison to Previous Determinations}\label{sec:metal}

Table \ref{table:table6} shows the metallicities determined for each of the target-cluster stars, 
the mean metallicities of each target-cluster, and the scatter in the target-cluster star metallicities.  
We report the error for each target-cluster as the error in the mean of the target-cluster giant stars: 
the error in the method (0.08 dex) divided by the 
square root of the number of giant stars measured in each target-cluster.  

Note that our method differs in a significant way from previous papers 
using spectrophotometric line indices to derive metallicities: we combine all of the 
measured indices at once using $\chi^2$ analysis, whereas others \citep[e.g.,][]{fj93,tff,friel02}
use each index individually to derive a metallicity for each star, then average 
the derived metallicities together.  Our 
analysis indicates that several of the indices provide no additional sensitivity to metallicity
for the stars in our sample, and we do not use them at all.  

Below we discuss the application of our fit to each target-cluster individually 
and compare the derived metallicity to existing values in the literature.
Most of the open clusters studied here do not have a large number of previous metallicity 
determinations.  

A new catalog of open clusters \citep[][]{dias} 
improves upon and replaces the \cite{lynga} database and includes metallicity 
estimates for all of our target-clusters.  
Although the on-line catalog does not include references for the metallicity 
estimates of the clusters, Dias (2005, private communication) has provided the 
references used in the catalog.
We cite the references given to us by Dias for each of the clusters in the discussion below.  

\subsubsection{NGC 1245}

NGC 1245 has a metallicity of [m/H]=$-0.14\pm$0.04 derived from seven measurements of six stars.  
The scatter in the derived metallicities is 0.09 dex, close to 
what we expect based on the error in the method.  
The identifiers used in Table \ref{table:table3} are taken from \cite{chincarini}.  
The cluster has a well-defined red clump and all target-cluster stars were taken from 
the clump.  We observed seven of the 27 clump giants from out instrumental CMD of NGC 1245.  
Unfortunately, WEBDA does not give membership probabilities for these stars.

NGC 1245 is studied extensively by \cite{paperi} (hereafter referred to as Paper I); 
new photometry is reported and 
fundamental properties of the cluster are derived, including a new metallicity value 
based on color-magnitude diagram profile fitting. 
It was selected as the first of several clusters for a targeted transiting planet search 
by the STEPSS program.  
Paper I reports a metallicity for NGC 1245 of [Fe/H]=$-0.05\pm$ 0.03 (statistical) 
$\pm$ 0.08 (systematic).  This value is derived from detailed isochrone fitting, 
using a $\chi^2$ fit to accurate photometric observations to derive physical 
parameters of the cluster and quantifying the systematic errors in the method.  
This method yields perhaps the most reliable derivation of cluster parameters
of any of those referred to below, with the possible exception of the 
high-resolution spectroscopy determinations.  

It should be noted that in \S 3.1 of Paper I a best-fit value of the total-to-selective
extinction $R_V$=2.3 is derived, yielding a metallicity of [Fe/H]=$-0.26$. 
This value was in fact not used in the determination as it was assumed 
to be too low and the fiducial value of $R_V$=3.2 is adopted.  
Our derived metallicity agrees equally well with both determinations of Paper I.
The assumed total-to-selective extinction may often be a large source of error in the determination of 
fundamental cluster parameters using isochrone fitting methods, 
since $R_V$ is not 3.2 uniformly throughout the Galaxy \citep[see, e.g.,][]{gould}.

\cite{weelee} also study the metallicity of NGC 1245 with Washington photometry.  
They derive a metallicity of [Fe/H]=$-0.04\pm$0.05. 
\cite{gratton} cite a value of [Fe/H]=$+0.1\pm$0.15, derived by applying 
corrections based on high-resolution spectroscopy to low-resolution metallicity determinations. 

The metallicity of NGC 1245 derived here, [m/H]=$-0.14\pm0.04$, is somewhat lower than previously published 
metallicities for this cluster.
In particular, our metallicity for NGC 1245 reproduces the value cited in 
Paper I to only $\sim$2 sigma.  We agree with the \cite{gratton} metallicity to 1.5 sigma and 
the \cite{weelee} metallicity to 1.6 sigma.  

\subsubsection{NGC 2099}

The metallicity of NGC 2099 is [m/H]=$+0.05\pm$0.05 and is derived from measurements 
of eight stars.
The metallicity determinations of the target-cluster stars in NGC 2099 
have a scatter of 0.14 dex, about twice what we expect.  
Target-cluster names in Table \ref{table:table3} are taken from \cite{vanz}.
We observed eight of the 23 clump giants in this cluster's instrumental CMD; 
all of the target-cluster giants in NGC 2099 are taken from the well-populated clump. 
The observed stars all had membership probabilities of greater than $75\%$, according to WEBDA.

NGC 2099 is a relatively unstudied cluster with few metallicity measurements.  
It is also the second cluster (after NGC 1245) to be observed by 
the STEPSS transiting planet search. 
\cite{twarog} are some of the few to have studied NGC 2099.  They use 
DDO photometry to derive [Fe/H]=+0.089$\pm$0.146.  
Our value of [m/H]=$+0.05\pm$0.05 is consistent with 
the metallicity cited by \cite{twarog} to 0.2 sigma.  

\subsubsection{NGC 2324}

We derive a metallicity for NGC 2324 of [m/H]=$-0.06\pm$0.04 from three measurements of four stars.
The scatter in metallicity determinations among the stars observed in 
NGC 2324, 0.07 dex, is close to the expected level given the error in the method.  
Identifiers for these target-cluster stars are from \cite{piatti}.
This cluster has a relatively well-defined clump; our target-cluster stars were 
selected from the seven giants in the clump.
WEBDA reports very low membership probabilities for two of these stars; 
0850 is given as $35\%$ and 1006 is $11\%$.  The third star, 1552, does not have a membership
probability reported in WEBDA.  These numbers are surprising, since our results 
have low scatter and the stars fall within the region of the clump in the CMD.  

\cite{friel02} studied NGC 2324 with the spectrophotometric line index method
used in this paper.  They report a metallicity for NGC 2324 of [Fe/H]=$-0.15\pm$0.16.  
Others have also studied this cluster: \cite{piatti} use their own reddening data with 
the Washington photometric data 
of \cite{geisler} to derive a metallicity of [Fe/H]=$-0.31\pm$0.04. 
\cite{geisler} themselves derive [Fe/H]$\sim-1$ for this cluster.
\cite{cameron} uses UV excess methods to derive [Fe/H]=-0.163.

We report [m/H]=$-0.06\pm$0.04 for NGC 2324. 
Our derived metallicity is consistent with the \cite{friel02} value to 0.5 sigma and 
with the \cite{piatti} value to 4.4 sigma.

\subsubsection{NGC 2539}

The metallicity of NGC 2539 is [m/H]=$-0.04\pm$0.03, derived from four stars.
The standard deviation of the metallicity estimates of the cluster stars observed in NGC 2539 is 0.05 dex.
We use the \cite{lapasset} identifiers for the target-cluster stars in this cluster.
Our instrumental CMD of NGC 2539 showed a small clump with only a few stars; we observed all 
four of these clump stars.
WEBDA does not report membership probabilities for this cluster, although due to 
the low scatter in metallicities we believe all of our target-cluster stars to 
in fact be members of the cluster.

\cite{cl86} derive a metallicity of [Fe/H]=$+0.2$ from DDO data; they also derive a 
value of [Fe/H]=$-0.2\pm$0.1 from iron lines.  
\cite{twarog} publish an [Fe/H] of +0.137$\pm$0.062 determined from DDO photometry.  

Our derived metallicity differs from that of \cite{twarog} by 2.6 sigma.  
We agree with the \cite{cl86} metallicity derived from iron lines 
to 1.5 sigma.

\subsubsection{NGC 2682}

We report a metallicity of [m/H]=$-0.05\pm$0.02 for NGC 2682 from five measurements of four cluster stars. 
The scatter is about half what we expect given the error in the method, 0.04 dex.   
\cite{fagerholm} provides the identifications for the stars in this target-cluster.
All of our target-cluster stars have greater than $93\%$ probabilities of being cluster members, 
according to WEBDA.
NGC 2682 has the most evolved giant branch of any of the target-clusters;
in this cluster we selected four stars from the instrumental CMD that were located 
along the giant branch.

NGC 2682 (M67) is one of the oldest known open clusters in the Galaxy; it is also 
by far the most-studied cluster in our sample, with many more metallicity 
determinations than discussed here.  
Previous studies agree that NGC 2682 has an approximately solar metallicity.  
\cite{cds90} uses high-resolution spectroscopy to obtain a metallicity 
for NGC 2682 of [Fe/H]=+0.039$\pm$0.09.  
\cite{hobbs} use high-resolution spectroscopy of weak iron lines to 
find [Fe/H]=$-0.04\pm0.12$.  
Both \cite{fj93} and \cite{friel02} derive the metallicity of this cluster from indices used in this paper; 
they obtain metallicities of [Fe/H]=$-0.09\pm$0.07 and [Fe/H]=$-0.15\pm$0.05, respectively.
\cite{burstein86} use a moderate resolution iron index to derive a metallicity
of [Fe/H]=$-0.1^{+0.12}_{-0.04}$.  
\cite{cameron} uses the cluster's UV excess to derive [Fe/H]=$-0.029$.
\cite{twarog} use DDO photometry to find [Fe/H]=+0.000$\pm$0.092.  

Our derived metallicity for NGC 2682 agrees well with 
most metallicities given in the literature.  
This metallicity agrees with \cite{cds90} to 0.96 sigma and with \cite{hobbs} to 0.1 sigma.  
Our determination agrees with other metallicities derived from spectrophotometric line index methods as well: 
we reproduce the \cite{fj93} value to 0.5 sigma and the \cite{friel02} value to 1.8 sigma.  
We agree with the lower error of \cite{burstein86} to 1.1 sigma; 
we agree with \cite{twarog} to 0.5 sigma.  

\subsubsection{NGC 6705}

The metallicity of NGC 6705 is [m/H]=+0.14$\pm$0.08, derived from four stars.
The stars have a scatter of 0.16 dex.  
We use the identifiers of \cite{mcnamara} for NGC 6705. These stars all had membership probabilities 
in WEBDA of $98\%$ or greater.
This cluster has a very sparse instrumental CMD; we observed four of the six giants in the well-defined clump.

\cite{cameron} also studied NGC 6705.  For this cluster they derive [Fe/H]=+0.070 with UV excess methods.  
\cite{twarog} give [Fe/H]=+0.136$\pm$0.086, derived by shifting spectroscopic measurements made by 
\cite{tff} onto a scale in common with other literature values.  
\cite{tff} derive [Fe/H]=$+0.21\pm$0.09 using spectrophotometric line indices.

We derive a metallicity for NGC 6705 of [m/H]=+0.14$\pm$0.08.  
This agrees very well with previous estimates; it is within 0.03 sigma of the metallicity derived by \cite{twarog}, 
and agrees with \cite{tff} to 0.6 sigma.  

\subsubsection{NGC 6819}

NGC 6819 has a metallicity of [m/H]=$+0.07\pm$0.12.
This cluster has the largest scatter in metallicity determinations among 
cluster stars in our sample (0.24 dex).  It is possible that some of the stars 
in our sample are not in fact cluster members. Unfortunately, WEBDA does not report 
membership probabilities for any of our stars in NGC 6819.
Furthermore, the four stars selected in this cluster lie slightly below the 
giant branch of this evolved cluster.  This may also account for some of the 
scatter in the derived metallicities of these stars.
The identifiers for this target-cluster are taken from \cite{auner}.

NGC 6819 is another old open cluster that has been studied more than the younger 
clusters.  The cluster's metallicity is derived in two papers using the index method:
both \cite{fj93} and \cite{tff} derive [Fe/H]=$+0.05\pm$0.11. 
\cite{bragaglia} use high-resolution spectroscopy to measure metal lines 
in three red clump stars in NGC 6819; 
they derive a metallicity of [Fe/H]=+0.09$\pm$0.03.  
\cite{twarog} give [Fe/H]=+0.074$\pm$0.123, again obtained by 
shifting spectroscopic measurements made by 
\cite{fj93} and \cite{tff} onto a scale in common with other literature values.

Our metallicity estimate for NGC 6819 is [m/H]=$+0.07\pm$0.12,  
in excellent agreement with existing literature values.
We agree with \cite{fj93} and \cite{tff} to 0.12 sigma, with 
\cite{twarog} to 0.02 sigma, and with \cite{bragaglia} to 0.16 sigma.   

%---------------------------------------------------------------------------
\section{Conclusions}\label{sec:concl}

We have derived metallicity estimates for seven Galactic open clusters.  
We use moderate-resolution spectroscopy to estimate metallicities of 
giants in these clusters, calibrating our results with 
high-resolution metallicity determinations of field giants.  
We reproduce the high-resolution determinations to 0.08 dex.  
Our derived metallicities generally agree with those found in the literature; 
more importantly, we provide new metallicity estimates for several clusters 
that are not well studied.  

This method is of interest because 
it requires little observing time, relatively small telescopes, and the results are arrived at 
quickly compared to more complicated methods (such as those used in Paper I).  
This method is also more robust than many photometric techniques, as 
it relies on quantitative spectroscopy and relative index measurements, not on 
the generally unpredictable photometricity of the night sky.  
We have demonstrated that it reproduces the metallicities of calibration stars of 
known metallicity well. 
The measured indices are easily reproducible on different telescopes 
and even, as evidenced by comparison of our indices to the published indices of \cite{friel87}, 
by different instruments and observers.  

The metallicity estimates given here will allow for selection of open clusters 
in which to conduct future transiting extra-solar planet searches.  

%---------------------------------------------------------------------------

\acknowledgements

The authors wish to thank W. Dias for making an updated version of part of his 
open cluster catalog available to us prior to publication.  
We have made extensive use of the WEBDA database 
as well as the VizieR Service.  

This work was supported by NASA grant NAGS-10678.
A. Gould was supported by grant AST 02-01266 from the NSF.  

%---------------------------------------------------------------------------

%---------------------------------------------------------------------------
%
% Figure Captions start here on a new page
%
%\newpage
%\section{Figure Captions}

%
%
%---------------------------------------------------------------------------
%
% Tables begin here, also on a new page
%
\begin{deluxetable}{cccc}
\tablecolumns{8}
\tablewidth{0pc}
\tablecaption{
Indices Measured
\label{table:table1}
}
\tablehead{
\colhead{Index} & \colhead{Central Bandpass}   & \colhead{Continuum Bandpass}    & \colhead{Species Measured}}
\startdata
Fe4065  & 4033.50 - 4085.00 & 4006.00 - 4038.00 & Fe I, Mn I, Sr II \\
        &                   & 4085.00 - 4115.00 & \\
CN4216  & 4144.00 - 4177.75 & 4082.00 - 4118.25 & CN ($B^2\Sigma^+  - X^2\Sigma^+$) \\
        &                   & 4246.00 - 4284.75 &            \\
Ca4226  & 4207.00 - 4245.00 & 4115.00 - 4165.00 & Ca I       \\
        &                   & 4320.00 - 4370.00 &            \\
G-band  & 4245.00 - 4320.00 & 4120.00 - 4200.00 & CH, Fe I   \\
        &                   & 4425.00 - 4520.00 &            \\
Fe4530  & 4520.00 - 4571.25 & 4498.75 - 4520.00 & Fe I, Ti I, Ti II \\
        &                   & 4606.00 - 4636.00 & \\
Fe4680  & 4636.00 - 4723.00 & 4606.00 - 4636.00 & Fe I, Cr I, Ni I, Mg I \\
        &                   & 4736.00 - 4773.00 & \\
Fe4920  & 4900.00 - 4940.00 & 4796.00 - 4841.00 & Fe I \\
        &                   & 4935.00 - 4975.00 & \\
Fe5011  & 4976.00 - 5051.00 & 4935.00 - 4975.00 & Fe I, Ti I, Fe II, Ni I \\
        &                   & 5051.00 - 5096.00 & \\
Mg      & 5130.00 - 5200.00 & 4935.00 - 4975.00 & Mg b + MbH \\
        &                   & 5303.00 - 5367.00 &            \\
Fe5270  & 5248.00 - 5286.75 & 5220.00 - 5250.00 & Fe I, Ca I \\
        &                   & 5288.00 - 5322.00 & \\
Fe5335  & 5314.75 - 5353.50 & 5307.25 - 5317.25 & Fe I, Cr I \\
        &                   & 5356.00 - 5364.75 & \\

\enddata
\end{deluxetable}
%
%
%\begin{landscape}
\begin{deluxetable}{lllcccccccccccc}
\rotate
\tabletypesize{\scriptsize}
\tablecolumns{15}
\tablewidth{0pc}
\tablecaption{
Metallicity indices for calibration stars
\label{table:table2}
}
\tablehead{
\colhead{Star Name} & \colhead{Date Observed} & \colhead{(B-V)$_0$\tablenotemark{a}} & \colhead{[Fe/H]\tablenotemark{b}}& \colhead{Ca4226} & \colhead{CN4216}& \colhead{Fe4065} & \colhead{Fe4530} & \colhead{Fe4680} & \colhead{Fe4920} & \colhead{Fe5011} & \colhead{Fe5270} & \colhead{Fe5335} & \colhead{Gband} & \colhead{Mg} }
\startdata
$\alpha$ UMa & Jan 31 & 1.061       & -0.19  & 0.783 & 0.982 & 0.822 & 0.883 & 0.811 & 0.838 & 0.818 & 0.825 & 0.828 & 0.914 & 0.889 \\
$\alpha$ UMa & Mar 25 & 1.061       & -0.19  & 0.793 & 0.996 & 0.819 & 0.886 & 0.819 & 0.816 & 0.806 & 0.829 & 0.833 & 0.969 & 0.936 \\
BD +3 2782 & Mar 29 & 1.092          & -2.0233 & 0.863 & 0.785 & 0.882 & 0.835 & 0.812 & 0.833 & 0.794 & 0.850 & 0.851 & 1.040 & 0.964 \\
$\beta$ Gem & Mar 31 & 0.991           & -0.0517 & 0.830 & 0.960 & 0.849 & 0.862 & 0.823 & 0.807 & 0.804 & 0.837 & 0.841 & 0.975 & 0.862 \\
$\epsilon$ Vir & Jan 31 & 0.934     & 0.148  & 0.777 & 0.954 & 0.832 & 0.855 & 0.819 & 0.847 & 0.819 & 0.833 & 0.824 & 0.900 & 0.887 \\
$\epsilon$ Vir & Mar 29 & 0.934     & 0.148  & 0.791 & 0.958 & 0.827 & 0.864 & 0.820 & 0.829 & 0.810 & 0.832 & 0.831 & 0.916 & 0.865 \\
$\epsilon$ Vir & Mar 31 & 0.934     & 0.148  & 0.801 & 0.960 & 0.830 & 0.863 & 0.820 & 0.810 & 0.807 & 0.827 & 0.828 & 0.927 & 0.845 \\
HD 108225 & Mar 22 & 0.955         & -0.11  & 0.806 & 0.936 & 0.830 & 0.853 & 0.819 & 0.818 & 0.805 & 0.835 & 0.836 & 0.930 & 0.875 \\
HD 108225 & Mar 31 & 0.955         & -0.11  & 0.814 & 0.933 & 0.841 & 0.857 & 0.822 & 0.816 & 0.805 & 0.831 & 0.829 & 0.949 & 0.893 \\
HD 111028 & Jan 31 & 0.989             & -0.4   & 0.817 & 0.849 & 0.844 & 0.848 & 0.813 & 0.831 & 0.800 & 0.834 & 0.828 & 1.008 & 0.917 \\
HD 111028 & Mar 23 & 0.989             & -0.4   & 0.822 & 0.860 & 0.846 & 0.846 & 0.812 & 0.812 & 0.789 & 0.834 & 0.842 & 1.004 & 0.923 \\
HD 111028 & Mar 24 & 0.989             & -0.4   & 0.819 & 0.859 & 0.838 & 0.846 & 0.813 & 0.824 & 0.794 & 0.833 & 0.838 & 0.997 & 0.938 \\
HD 117876 & Jan 31 & 0.969         & -0.5   & 0.790 & 0.839 & 0.815 & 0.846 & 0.791 & 0.821 & 0.797 & 0.816 & 0.820 & 0.969 & 0.873 \\
HD 117876 & Mar 22 & 0.969         & -0.5   & 0.801 & 0.851 & 0.811 & 0.850 & 0.795 & 0.800 & 0.793 & 0.816 & 0.818 & 0.975 & 0.880 \\
HD 117876 & Mar 24 & 0.969         & -0.5   & 0.796 & 0.849 & 0.813 & 0.848 & 0.792 & 0.814 & 0.794 & 0.813 & 0.815 & 0.972 & 0.884 \\
HD 122563 & Mar 22 & 0.853         & -2.683 & 0.770 & 0.752 & 0.737 & 0.779 & 0.753 & 0.768 & 0.749 & 0.760 & 0.764 & 0.862 & 0.732 \\
HD 122563 & Mar 31 & 0.853         & -2.683 & 0.786 & 0.756 & 0.744 & 0.785 & 0.756 & 0.766 & 0.748 & 0.762 & 0.765 & 0.881 & 0.763 \\
HD 129312 & Mar 23 & 0.992         & -0.31  & 0.785 & 0.981 & 0.826 & 0.867 & 0.819 & 0.818 & 0.812 & 0.834 & 0.831 & 0.910 & 0.908 \\
HD 130952 & Jan 31 & 0.988         & -0.395 & 0.799 & 0.861 & 0.832 & 0.857 & 0.802 & 0.852 & 0.811 & 0.809 & 0.825 & 1.034 & 0.917 \\
HD 130952 & Mar 24 & 0.988         & -0.395 & 0.798 & 0.862 & 0.816 & 0.849 & 0.805 & 0.809 & 0.793 & 0.821 & 0.838 & 1.005 & 0.892 \\
HD 135722 & Mar 24 & 0.961         & -0.47  & 0.799 & 0.829 & 0.824 & 0.851 & 0.795 & 0.816 & 0.797 & 0.812 & 0.813 & 0.991 & 0.863 \\
HD 138905 & Mar 24 & 1.007         & -0.41  & 0.801 & 0.874 & 0.831 & 0.853 & 0.797 & 0.819 & 0.795 & 0.818 & 0.821 & 0.992 & 0.877 \\
HD 138905 & Mar 31 & 1.007         & -0.41  & 0.828 & 0.886 & 0.831 & 0.860 & 0.796 & 0.804 & 0.794 & 0.821 & 0.821 & 1.022 & 0.844 \\
HD 150275 & Mar 24 & 0.995         & -0.5   & 0.818 & 0.791 & 0.813 & 0.850 & 0.787 & 0.801 & 0.790 & 0.809 & 0.810 & 1.043 & 0.908 \\
HD 150275 & Mar 31 & 0.995         & -0.5   & 0.835 & 0.802 & 0.818 & 0.859 & 0.791 & 0.782 & 0.786 & 0.816 & 0.820 & 1.077 & 0.853 \\
HD 168322 & Mar 24 & 0.977         & -0.4   & 0.800 & 0.815 & 0.810 & 0.848 & 0.799 & 0.814 & 0.798 & 0.809 & 0.811 & 1.012 & 0.887 \\
HD 198149 & Apr 01 & 0.912         & -0.41  & 0.810 & 0.833 & 0.832 & 0.840 & 0.804 & 0.818 & 0.787 & 0.820 & 0.822 & 1.011 & 0.890 \\
HD 199191 & Apr 01 & 0.960         & -0.7   & 0.806 & 0.776 & 0.810 & 0.835 & 0.786 & 0.801 & 0.783 & 0.805 & 0.808 & 1.028 & 0.882 \\
HD 25975 & Mar 30 & 0.943         & -0.2   & 0.823 & 0.895 & 0.844 & 0.847 & 0.820 & 0.834 & 0.798 & 0.842 & 0.840 & 0.989 & 0.991 \\
HD 27371 & Mar 24 & 0.981          & 0.1128 & 0.788 & 1.021 & 0.832 & 0.859 & 0.830 & 0.828 & 0.811 & 0.835 & 0.837 & 0.914 & 0.869 \\
HD 27971 & Mar 31 & 0.986          & -0.08  & 0.813 & 0.971 & 0.842 & 0.863 & 0.827 & 0.818 & 0.807 & 0.832 & 0.835 & 0.957 & 0.860 \\
HD 37160 & Mar 30 & 0.951          & -0.5533 & 0.816 & 0.792 & 0.811 & 0.834 & 0.770 & 0.816 & 0.779 & 0.814 & 0.820 & 1.052 & 0.891 \\
HD 37160 & Mar 31 & 0.951          & -0.5533 & 0.823 & 0.793 & 0.814 & 0.844 & 0.790 & 0.785 & 0.781 & 0.812 & 0.817 & 1.071 & 0.856 \\
HD 40460 & Mar 30 & 1.022          & -0.5   & 0.814 & 0.902 & 0.834 & 0.855 & 0.810 & 0.818 & 0.793 & 0.830 & 0.836 & 1.002 & 0.928 \\
HD 41636 & Mar 30 & 1.044          & -0.3   & 0.801 & 0.914 & 0.825 & 0.845 & 0.814 & 0.824 & 0.798 & 0.830 & 0.829 & 0.990 & 0.975 \\
HD 43039 & Mar 30 & 1.021          & -0.415 & 0.819 & 0.886 & 0.831 & 0.850 & 0.804 & 0.819 & 0.796 & 0.829 & 0.831 & 1.006 & 0.937 \\
HD 43039 & Mar 31 & 1.021          & -0.415 & 0.821 & 0.887 & 0.831 & 0.860 & 0.809 & 0.802 & 0.796 & 0.824 & 0.829 & 1.025 & 0.883 \\
HD 46480 & Jan 31 & 0.899          & -0.5   & 0.784 & 0.759 & 0.810 & 0.832 & 0.784 & 0.832 & 0.792 & 0.799 & 0.807 & 1.003 & 0.932 \\
HD 46480 & Mar 31 & 0.899          & -0.5   & 0.817 & 0.776 & 0.813 & 0.840 & 0.792 & 0.791 & 0.782 & 0.803 & 0.808 & 1.051 & 0.863 \\
HD 62345 & Mar 29 & 0.932          & -0.18  & 0.789 & 0.947 & 0.828 & 0.855 & 0.813 & 0.824 & 0.804 & 0.829 & 0.828 & 0.931 & 0.867 \\
HD 63410 & Jan 31 & 0.961          & -0.5   & 0.787 & 0.842 & 0.815 & 0.849 & 0.803 & 0.829 & 0.805 & 0.813 & 0.814 & 0.988 & 0.881 \\
HD 63410 & Mar 30 & 0.961          & -0.5   & 0.806 & 0.851 & 0.816 & 0.840 & 0.797 & 0.820 & 0.790 & 0.822 & 0.824 & 1.003 & 0.903 \\
HD 73710 & Mar 31 & 1.020          & 0.11   & 0.810 & 1.065 & 0.843 & 0.887 & 0.841 & 0.811 & 0.812 & 0.841 & 0.845 & 0.932 & 0.866 \\
HD 81192 & Jan 31 & 0.933          & -0.65  & 0.795 & 0.739 & 0.804 & 0.834 & 0.777 & 0.820 & 0.787 & 0.802 & 0.795 & 1.023 & 0.866 \\
HD 81192 & Mar 23 & 0.933          & -0.65  & 0.803 & 0.751 & 0.802 & 0.837 & 0.778 & 0.795 & 0.781 & 0.811 & 0.812 & 1.031 & 0.913 \\
HD 81192 & Mar 23 & 0.933          & -0.65  & 0.813 & 0.752 & 0.799 & 0.832 & 0.798 & 0.770 & 0.788 & 0.774 & 0.773 & 1.044 & 0.487 \\
HD 85773 & Mar 29 & 1.078          & -2.31  & 0.773 & 0.807 & 0.762 & 0.797 & 0.764 & 0.778 & 0.757 & 0.777 & 0.778 & 0.904 & 0.771 \\
HD 88609 & Mar 23 & 0.894          & -2.845 & 0.769 & 0.768 & 0.760 & 0.778 & 0.749 & 0.760 & 0.746 & 0.765 & 0.762 & 0.847 & 0.777 \\
HD 91612 & Mar 23 & 0.921          & -0.175 & 0.795 & 0.851 & 0.826 & 0.850 & 0.802 & 0.812 & 0.798 & 0.818 & 0.813 & 0.958 & 0.825 \\
HD 96436 & Mar 23 & 0.955          & -0.5   & 0.801 & 0.812 & 0.814 & 0.841 & 0.796 & 0.803 & 0.786 & 0.819 & 0.828 & 1.021 & 0.906 \\
NGC 2682 0135 & Mar 30 & 1.01      & -0.09  & 0.829 & 0.966 & 0.847 & 0.867 & 0.823 & 0.818 & 0.797 & 0.840 & 0.848 & 0.962 & 0.906 \\
NGC 2682 0141 & Mar 30 & 1.06      & 0.0    & 0.814 & 1.009 & 0.847 & 0.863 & 0.830 & 0.816 & 0.804 & 0.839 & 0.839 & 0.942 & 0.889 \\
NGC 2682 0164 & Mar 31 & 1.07      & 0.0    & 0.837 & 1.032 & 0.853 & 0.874 & 0.828 & 0.801 & 0.806 & 0.839 & 0.845 & 1.002 & 0.885 \\
NGC 2682 0224 & Mar 31 & 1.08      & -0.195 & 0.858 & 1.064 & 0.839 & 0.878 & 0.816 & 0.802 & 0.804 & 0.839 & 0.850 & 0.948 & 0.898 \\
NGC 2682 0084 & Mar 30 & 1.06      & -0.035 & 0.814 & 0.998 & 0.845 & 0.871 & 0.827 & 0.816 & 0.803 & 0.835 & 0.845 & 0.960 & 0.877 \\
{\rm o} Vir & Mar 29 & 0.967        & -0.29  & 0.783 & 0.849 & 0.834 & 0.851 & 0.802 & 0.817 & 0.789 & 0.807 & 0.808 & 0.999 & 0.839 \\
\enddata 
\tablenotetext{a}{Colors taken from the Hipparcos Catalog \citep[][]{esa97}}
\tablenotetext{b}{Metallicities derived from \cite{cds01}, as described in \S \ref{sec:analysis}}
\end{deluxetable}
%\end{landscape}
%
%
%\begin{landscape}
\begin{deluxetable}{lllccccccccccc}
\rotate
\tabletypesize{\scriptsize}
\tablecolumns{14}
\tablewidth{0pc}
\tablecaption{
Metallicity indices for target-cluster stars
\label{table:table3}
}
\tablehead{
\colhead{Star Name\tablenotemark{a}} & \colhead{Date Observed} & \colhead{(B-V)$_0$\tablenotemark{b}} & \colhead{Ca4226} & \colhead{CN4216}& \colhead{Fe4065} & \colhead{Fe4530} & \colhead{Fe4680} & \colhead{Fe4920} & \colhead{Fe5011} & \colhead{Fe5270} & \colhead{Fe5335} & \colhead{Gband} & \colhead{Mg} }
\startdata
NGC 1245 0010 & Jan 31 & 0.878 & 0.779 & 0.836 & 0.829 & 0.851 & 0.799 & 0.829 & 0.809 & 0.816 & 0.811 & 0.959 & 0.859 \\
NGC 1245 0014 & Jan 31 & 0.914 & 0.778 & 0.814 & 0.825 & 0.842 & 0.804 & 0.826 & 0.815 & 0.811 & 0.807 & 0.962 & 0.874 \\
NGC 1245 0132 & Mar 30 & 0.899 & 0.819 & 0.852 & 0.819 & 0.845 & 0.805 & 0.809 & 0.794 & 0.821 & 0.818 & 0.980 & 0.879 \\
NGC 1245 0148 & Jan 31 & 0.918 & 0.768 & 0.851 & 0.805 & 0.849 & 0.800 & 0.833 & 0.812 & 0.818 & 0.813 & 0.956 & 0.881 \\
NGC 1245 0162 & Mar 31 & 0.897 & 0.814 & 0.874 & 0.777 & 0.822 & 0.800 & 0.790 & 0.798 & 0.783 & 0.756 & 0.973 & 0.834 \\
NGC 1245 0395 & Mar 24 & 0.896 & 0.773 & 0.849 & 0.816 & 0.835 & 0.808 & 0.822 & 0.838 & 0.822 & 0.810 & 0.962 & 0.893 \\
NGC 2099 0004 & Jan 31 & 0.88 & 0.832 & 1.051 & 0.864 & 0.890 & 0.830 & 0.832 & 0.823 & 0.845 & 0.843 & 0.960 & 0.954 \\
NGC 2099 0031 & Jan 31 & 0.90 & 0.779 & 0.923 & 0.831 & 0.858 & 0.810 & 0.852 & 0.820 & 0.821 & 0.809 & 0.934 & 0.890 \\
NGC 2099 0049 & Jan 31 & 0.90 & 0.717 & 0.814 & 0.765 & 0.839 & 0.796 & 0.835 & 0.812 & 0.809 & 0.797 & 0.895 & 0.879 \\
NGC 2099 0064 & Mar 25 & 0.96 & 0.810 & 1.021 & 0.841 & 0.875 & 0.818 & 0.819 & 0.815 & 0.840 & 0.831 & 0.969 & 0.939 \\
NGC 2099 0148 & Mar 31 & 0.98 & 0.806 & 1.009 & 0.834 & 0.878 & 0.822 & 0.805 & 0.812 & 0.829 & 0.822 & 0.960 & 0.847 \\
NGC 2099 0439 & Mar 25 & 0.90 & 0.795 & 0.931 & 0.826 & 0.862 & 0.809 & 0.814 & 0.801 & 0.827 & 0.825 & 0.954 & 0.859 \\
NGC 2099 0454 & Mar 25 & 0.94 & 0.786 & 1.009 & 0.826 & 0.878 & 0.820 & 0.818 & 0.813 & 0.839 & 0.828 & 0.965 & 0.921 \\
NGC 2099 0716 & Mar 25 & 0.87 & 0.789 & 0.913 & 0.815 & 0.859 & 0.803 & 0.816 & 0.801 & 0.826 & 0.816 & 0.959 & 0.890 \\
NGC 2324 0850 & Mar 30 & 0.75 & 0.791 & 0.840 & 0.768 & 0.856 & 0.789 & 0.792 & 0.799 & 0.806 & 0.809 & 0.957 & 0.782 \\
NGC 2324 0850 & Mar 31 & 0.75 & 0.774 & 0.813 & 0.763 & 0.851 & 0.798 & 0.794 & 0.798 & 0.809 & 0.813 & 0.965 & 0.799 \\
NGC 2324 1006 & Mar 30 & 0.77 & 0.754 & 0.826 & 0.763 & 0.847 & 0.792 & 0.794 & 0.798 & 0.821 & 0.807 & 0.941 & 0.791 \\
NGC 2324 1552 & Mar 31 & 0.79 & 0.800 & 0.858 & 0.812 & 0.862 & 0.801 & 0.783 & 0.799 & 0.810 & 0.807 & 0.999 & 0.808 \\
NGC 2539 0114 & Mar 29 & 0.906 & 0.813 & 0.933 & 0.815 & 0.857 & 0.812 & 0.819 & 0.809 & 0.845 & 0.832 & 0.970 & 0.846 \\
NGC 2539 0229 & Mar 29 & 0.926 & 0.804 & 0.957 & 0.828 & 0.857 & 0.810 & 0.822 & 0.804 & 0.832 & 0.830 & 0.955 & 0.862 \\
NGC 2539 0246 & Mar 29 & 0.899 & 0.809 & 0.927 & 0.829 & 0.859 & 0.812 & 0.821 & 0.805 & 0.831 & 0.831 & 0.952 & 0.866 \\
NGC 2539 0251 & Mar 29 & 0.903 & 0.800 & 0.907 & 0.815 & 0.851 & 0.803 & 0.823 & 0.800 & 0.825 & 0.829 & 0.945 & 0.862 \\
NGC 2682 0020 & Mar 25 & 0.801 & 0.794 & 0.786 & 0.828 & 0.829 & 0.802 & 0.822 & 0.779 & 0.823 & 0.819 & 1.038 & 0.932 \\
NGC 2682 0037 & Mar 24 & 0.891 & 0.820 & 0.830 & 0.858 & 0.838 & 0.823 & 0.827 & 0.793 & 0.835 & 0.836 & 1.025 & 0.938 \\
NGC 2682 0166 & Mar 22 & 0.867 & 0.810 & 0.817 & 0.846 & 0.832 & 0.823 & 0.813 & 0.792 & 0.834 & 0.832 & 1.025 & 0.928 \\
NGC 2682 0166 & Mar 24 & 0.867 & 0.810 & 0.816 & 0.830 & 0.830 & 0.821 & 0.825 & 0.789 & 0.829 & 0.831 & 1.014 & 0.920 \\
NGC 2682 4018 & Mar 22 & 0.827 & 0.799 & 0.753 & 0.829 & 0.828 & 0.815 & 0.812 & 0.790 & 0.830 & 0.829 & 1.017 & 0.895 \\
NGC 6705 0686 & Mar 25 & 1.05 & 0.822 & 1.111 & 0.849 & 0.884 & 0.846 & 0.824 & 0.813 & 0.857 & 0.863 & 0.934 & 0.972 \\
NGC 6705 0916 & Mar 25 & 1.00 & 0.796 & 1.169 & 0.838 & 0.880 & 0.849 & 0.828 & 0.818 & 0.861 & 0.859 & 0.906 & 0.954 \\
NGC 6705 1145 & Mar 29 & 0.98 & 0.850 & 1.107 & 0.850 & 0.886 & 0.839 & 0.818 & 0.812 & 0.846 & 0.845 & 0.953 & 0.913 \\
NGC 6705 1223 & Mar 29 & 0.75 & 0.711 & 0.889 & 0.694 & 0.862 & 0.824 & 0.813 & 0.811 & 0.833 & 0.829 & 0.882 & 0.882 \\
NGC 6819 0281 & Mar 30 & 0.98 & 0.900 & 0.893 & 0.848 & 0.884 & 0.817 & 0.802 & 0.792 & 0.857 & 0.868 & 0.987 & 0.993 \\
NGC 6819 0287 & Mar 29 & 0.87 & 0.833 & 0.981 & 0.877 & 0.844 & 0.829 & 0.796 & 0.812 & 0.842 & 0.853 & 1.013 & 0.902 \\
NGC 6819 0339 & Mar 30 & 0.86 & 0.803 & 0.836 & 0.852 & 0.849 & 0.832 & 0.823 & 0.827 & 0.817 & 0.844 & 1.022 & 0.868 \\
NGC 6819 1306 & Mar 31 & 0.93 & 0.809 & 0.909 & 0.870 & 0.842 & 0.817 & 0.806 & 0.806 & 0.817 & 0.811 & 0.963 & 0.889 \\
\enddata
\tablenotetext{a}{Identifications as given in the WEBDA database}
\tablenotetext{b}{Sources for photometry and reddening: \\
NGC 1245: (B-V) and E(B-V)=0.25 from \cite{paperi} \\
NGC 2099: (B-V) from \cite{kiss} (via WEBDA), E(B-V)=0.29 \citep[][]{mermiliod96} \\
NGC 2324: (B-V) from \cite{kyeong} (via WEBDA), E(B-V)=0.25 \citep[][]{piatti}\\
NGC 2539: (B-V) and E(B-V)=0.06 from \cite{lapasset}\\
NGC 2682: (B-V) and E(B-V)=0.05 from \cite{mont}\\
NGC 6705: (B-V) from \cite{stetson} (via WEBDA), E(B-V)=0.43 \citep[][]{sung}\\
NGC 6819: (B-V) and E(B-V)=0.16 from \citep[][]{rosvick} ((B-V) via WEBDA)\\
}
\end{deluxetable}
%\end{landscape}
%
% 
\begin{deluxetable}{lll}
\tablecolumns{3}
\tablewidth{0pc}
\tablecaption{Comparison of Indices}
\tablehead{
\colhead{Subset of Data} & \colhead{Average of Difference} & \colhead{Standard Deviation of Difference} \\
\colhead{} & \colhead{for All Indices} & \colhead{for All Indices}
\label{table:table4}
}
\startdata
1.3 m \& 2.4 m & 0.00033 & 0.0335 \\
1.3 m \& 1.3 m & 0.0028 & 0.0384 \\
All stars & 0.0011 & 0.0287 \\
\enddata
\end{deluxetable}
\begin{deluxetable}{ll}
\tablecolumns{2}
\tablewidth{0pc}
\tablecaption{
Coefficients of the Fit
\label{table:table5}
}
\tablehead{
\colhead{Index} & \colhead{Coefficient}}
\startdata
Const   & -3.180415 \\
Fe4065  & 4.398434 \\
CN 4216 & 1.105911 \\
Ca4226  & -2.480848 \\
G-band  & -0.725162 \\
Fe4530  & 0 \\
Fe4680  & 2.835423 \\
Fe4920  & -1.620898 \\
Fe5011  & 0 \\
Mg      & 0 \\
Fe5270  & 0 \\
Fe5335  & 0 \\
\enddata
\end{deluxetable}
\begin{deluxetable}{lccccc} 
\tablecolumns{6} 
\tablewidth{0pc} 
\tablecaption{
Derived Cluster Metallicities
\label{table:table6}
}
\tablehead{ 
\colhead{Cluster} & \colhead{Star} & \colhead{(B-V)$_0$} & \colhead{Clump Star} & \colhead{Cluster} & \colhead{Cluster} \\
\colhead{Name} & \colhead{Name} & \colhead{} & \colhead{[m/H]} & \colhead{[m/H]} & \colhead{$\sigma$}
}
\startdata 

\cline{1-6}
NGC 1245 & & & & -0.14 & 0.09 \\ 
\cline{1-6}
& 0010 & 1.128 & -0.043 & & \\
& 0014 & 1.164 & -0.120 & & \\
& 0132 & 1.149 & -0.163 & & \\
& 0148 & 1.168 & -0.163 & & \\
& 0162 & 1.147 & -0.287 & & \\
& 0395 & 1.146 & -0.057 & & \\

\cline{1-6}
NGC 2099 & & & & +0.05 & 0.14 \\
\cline{1-6}
& 0004 & 0.87 & +0.297 & & \\
& 0031 & 0.89 & +0.040 & & \\
& 0049 & 0.89 & -0.198 & & \\
& 0064 & 0.95 & +0.080 & & \\
& 0148 & 0.97 & +0.059 & & \\
& 0439 & 0.89 & +0.034 & & \\
& 0454 & 0.93 & +0.102 & & \\
& 0716 & 0.86 & +0.002 & & \\

\cline{1-6}
NGC 2324 & & & & -0.06 & 0.07 \\
\cline{1-6}
& 0850 & 0.87 & -0.065 & & \\
& 0850 & 0.87 & -0.115 & & \\
& 1006 & 0.89 & -0.102 & & \\
& 1552 & 0.91 & +0.040 & & \\

\cline{1-6}
NGC 2539 & & & & -0.04 & 0.05 \\
\cline{1-6}
& 0114 & 0.886 & -0.078 & & \\
& 0229 & 0.906 & +0.003  & & \\
& 0246 & 0.879 & +0.009  & & \\
& 0251 & 0.883 & -0.084  & & \\

\cline{1-6}
NGC 2682 & & & & -0.05  & 0.04  \\
\cline{1-6}
& 0020 & 0.791 & -0.061  & & \\
& 0037 & 0.881 & -0.019  & & \\
& 0166 & 0.857 & -0.001  & & \\
& 0166 & 0.857 & -0.091  & & \\
& 4018 & 0.817 & -0.074  & & \\

\cline{1-6}
NGC 6705 & & & & +0.14 & 0.16 \\
\cline{1-6}
& 0686 & 1.05 & +0.144  & & \\
& 0916 & 1.00 & +0.324  & & \\
& 1145 & 0.98 & +0.157  & & \\
& 1223 & 0.75 & -0.065  & & \\

\cline{1-6}
NGC 6819 & & & & +0.07 & 0.24 \\
\cline{1-6}
& 0281 & 0.90 & -0.273 & & \\
& 0287 & 0.79 & +0.308 & & \\
& 0339 & 0.78 & +0.082 & & \\
& 1306 & 0.85 & +0.152 & & \\

\enddata

\end{deluxetable}
\end{document}